\documentclass[twocolumn]{emulateapj}
\usepackage{natbib}

\usepackage{epsfig}

\providecommand{\eprint}[1]{\href{http://arxiv.org/abs/#1}{#1}}

\providecommand{\adsurl}[1]{\href{#1}{}}

\begin{document}

\newcommand{\slnr}{\sigma_{\ln R}}
\newcommand{\phii}{P_{\delta_{\rm HII}}}
\newcommand{\wrkone}{\langle W_R^2(k)\rangle}
\newcommand{\wrktwo}{\langle W_R(k)\rangle}
\newcommand{\wrone}{\langle W_R^2\rangle}
\newcommand{\wrtwo}{\langle W_R\rangle}
\newcommand{\dhii}{\delta_{\rm HII}}
\newcommand{\pshot}{P^{1b}_{\delta x\delta x}}
\newcommand{\pconv}{P^{1b}_{\delta x\delta~\delta x\delta}}
\newcommand{\clb}{C_{\ell}^B}
\newcommand{\cclb}{\mathcal{C}_{\ell}^B}
\newcommand{\lpeak}{\ell_{{\rm peak}}}
\newcommand{\clpeak}{\left(\frac{\ell(\ell+1)}{2\pi}C_{\ell}^B\right)_{{\rm peak}}}
\newcommand{\clpeakt}{(\ell(\ell+1)C_{\ell}^B/2\pi)_{{\rm peak}}}
\newcommand{\cclbpeak}{(\cclb)_{\rm peak}}
\newcommand{\erf}{{\rm erf}}

\title{The Maximum B-mode Polarization of the Cosmic Microwave Background from Inhomogeneous Reionization}

\author{Michael J. Mortonson$^{1,2}$ and Wayne Hu$^{1,3}$}
\affil{$^{1}$Kavli Institute for Cosmological Physics, 
Enrico Fermi Institute, University of Chicago, Chicago, IL 60637\\
$^{2}$Department of Physics,  University of Chicago, Chicago, IL 60637\\
$^{3}$Department of Astronomy and Astrophysics, University of Chicago, Chicago, IL 60637
}
\begin{abstract}
We compute the $B$-mode polarization power spectrum of the CMB from an epoch of 
inhomogeneous reionization, using a simple model in which \ion{H}{2} regions are 
represented by ionized spherical bubbles with a log normal distribution of sizes whose
clustering properties are determined by large-scale structure. 
Both the global ionization fraction and the characteristic radius of 
\ion{H}{2} regions are allowed to be free functions of redshift.  
Models that would produce substantial contamination to degree scale
gravitational wave $B$-mode measurements have power that is dominated by the shot noise
of the bubbles.  Rare bubbles of $\gtrsim 100$~Mpc at $z>20$ can produce signals
that in fact exceed the $B$-modes from gravitational lensing and are comparable to the maximal
allowed signal of gravitational waves ($\sim 0.1~\mu$K) while
still being consistent with global constraints on the total optical depth.  Even bubbles down to $20$ Mpc at $z\sim 15$, or $40$ Mpc at $z\sim 10$
can be relevant ($0.01~\mu$K)  once the lensing signal is removed either statistically or directly. 
However, currently favored theoretical models that have ionization bubbles that only grow to such sizes
at the very end of a fairly prompt and late reionization produce signals which are at most at these
levels. 
\end{abstract}

\section{Introduction} \label{intro}

The polarization of the cosmic microwave background (CMB) can be decomposed into 
two components, \emph{E}-modes and \emph{B}-modes, which are geometrically
distinct but non-local 
combinations of the Stokes parameters $Q$ and $U$.  The two modes are distinguished
by the relationship between the direction of polarization and its spatial gradients.
In linear perturbation theory,
both scalar and tensor 
perturbations contribute to the $E$-modes, but scalar perturbations do not 
produce $B$-modes~\citep{kamionkowski_etal97,seljak_zaldarriaga97}. 
This fact makes observation of the CMB 
$B$-modes an ideal way to detect the tensor perturbations from gravitational
waves generated during
inflation.    
Detection of gravitational waves  would be strong evidence in favor of inflation, 
would determine the energy scale of inflation, 
and would help constrain models of inflation~\citep[see e.g.][for recent assessments]{gold_albrecht03,verde_etal06}.   This detection is the goal of a concerted experimental
effort currently underway \citep{TaskForce}.

Beyond linear perturbations, scalar density fluctuations can generate \emph{B}-modes.
For example, gravitational lensing spatially distorts the polarization field in a
manner that is independent of its orientation and hence generates \emph{B}-modes
from the intrinsic $E$-modes from recombination~\citep{zaldarriaga_seljak98}.   
Lensing places a fundamental limitation on the detection of inflationary gravitational
waves that is not far in energy scale from current limits \citep[e.g.,][]{page_etal06}
 if only polarization power spectra are measured.  Fortunately, with high signal-to-noise,
 high resolution polarization maps, de-lensing of the polarization field is in principle possible
 \citep{HuOka01,HirSel02}
 and may improve limits by an order of magnitude or more in polarization power
 \citep{KnoSon02,KesCooKam02,seljak_hirata04}.
 In addition, predictions for the gravitational lensing \emph{B}-modes
 are highly accurate and hence their contamination to a power spectrum measurement
 may be subtracted to within a few percent for a cosmic variance limited experiment. 
 
 More worrying would be cosmological \emph{B}-mode sources that cannot be accurately
 modelled theoretically.  As pointed out in \cite{hu00}, inhomogeneities in the free electron
 density during reionization will generate \emph{B}-modes through Thomson scattering.
Fluctuations in the local ionization fraction during an extended epoch of reionization are in 
fact expected.  
Current observations show evidence for an extended epoch of reionization that 
ends at redshift $z\sim 6$ and begins at $z\gtrsim 10$~\citep{fan_etal06}. 
Because reionization 
is a gradual process and the sources of ionizing radiation are 
clustered, the distribution of neutral and ionized hydrogen during the epoch
of reionization is expected to be highly inhomogeneous~\citep[e.g.,][]
{barkana_loeb01}.  

Previous studies of \emph{B}-modes generated by inhomogeneous (or ``patchy'') 
reionization have shown that 
they are negligible compared with  \emph{B}-modes from lensing at
scales relevant for gravitational waves 
if the ionized regions are small and uncorrelated
with each other and reionization is prompt~\citep{hu00,weller99}. 
Correlations between ionized regions can boost the signal but not substantially~\citep{knox_etal98,santos_etal03,baumann_etal03}.
However, because reionization is a highly uncertain process, it has not been possible
to  rule it out as a source of significant \emph{B}-modes when these conditions are violated.

In this paper, we take a phenomenological approach to the study of \emph{B}-mode
generation from inhomogeneous reionization to bound the possible contamination
to gravitational wave studies.  
We limit our considerations to models that satisfy observational bounds on the
total optical depth $\tau_* = 0.09 \pm 0.03$~\citep{spergel_etal06}.  We fix the other cosmological
parameters to also be consistent  with the 3-year 
WMAP results: Hubble constant $H_0=100 h$ 
km s$^{-1}$ Mpc$^{-1}$ with $h=0.73$, matter density $\Omega_mh^2=0.128$, 
baryon density $\Omega_bh^2=0.0223$, flat spatial geometry 
$\Omega_K=0$, CMB temperature $T_{{\rm CMB}}=2.725$ K, 
primordial helium fraction $Y_p=0.24$, 
scalar index $n_S=0.958$, and amplitude $\sigma_8=0.76$. 

Our approach should be contrasted with studies of physically motivated 
models of reionization~\citep{liu_etal01,zahn_etal05}.   
 The advantage 
of our approach is its ability to 
describe a wide variety of reionization scenarios with relatively few 
phenomenological parameters.   The parameters that are required
to generate substantial \emph{B}-modes can then be compared with predictions
from physical models.

The outline of the paper is as follows.  In Section~\ref{model}, we introduce a phenomenological
model of inhomogeneous reionization that incorporates several free functions of redshift.   We parameterize
these functions in Section~\ref{parameterization} in a manner which highlights the phenomenology of
the \emph{B}-modes discussed in  Section~\ref{bmodes}.  We discuss observational  and
theoretical bounds on the parameter space in Section~\ref{bounds} and conclude in  Section~\ref{discussion}.

\section{Inhomogeneous Reionization Models} \label{model}

Reionization generates polarization by Thomson scattering of anisotropic
radiation coming from recombination.  If the ionized hydrogen is homogeneous,
the scalar quadrupole temperature anisotropy creates only $E$-mode polarization.  
Inhomogeneities break this symmetry and generate $B$-mode
polarization.  
Under the Limber approximation, valid for multipoles $\ell \gg 10$, 
the \emph{B}-mode power spectrum is related
to the power spectrum of the ionized hydrogen 
$\phii(k,z)$ as~\citep{hu00}
\begin{eqnarray}
C_{\ell}^B&=&\frac{3\sigma_T^2n_{p,0}^2}{100}\int_{z_f}^{z_i} 
\frac{dz(1+z)^4}{H(z)D_A^2(z)}e^{-2\tau(z)} Q_{{\rm rms}}^2(z) \nonumber\\
&&\quad \times
\phii(k=\ell/D_{A},z), 
\label{clrei}
\end{eqnarray}
where $\sigma_T$ is the Thomson scattering cross section, $n_{p,0}$ is the 
present number density of protons, $z_i$ and $z_f$ are the redshifts at the 
beginning and end of the epoch of reionization, $H(z)$ is the Hubble 
parameter, $D_A(z)$ is the angular diameter distance, 
$\tau(z)$ is the optical depth out to redshift $z$, and  
$Q_{{\rm rms}}^2(z)$ is the variance of the quadrupole. 
We define the ionization fraction to equal 1 when both hydrogen and 
helium are fully ionized, so $n_{p,0}=(1-Y_p/2)n_{b,0}$. 
In the fiducial model $Q_{\rm rms} \approx 18~\mu$K at the redshifts of interest.
We often use the notation
\begin{eqnarray}
\cclb\equiv\ell(\ell+1)\clb/2\pi \,,
\end{eqnarray}
for the power per logarithmic interval in multipole.

To evaluate equation~(\ref{clrei}), we need a model for inhomogeneous 
reionization. Because we are interested in exploring what types of ionization
histories generate maximal amounts of $B$-modes, we adopt the
phenomenological parameterization of~\citet{wang_hu06} instead of a more
physically motivated one \citep[e.g.,][]{furlanetto_etal04}.  

In this parameterization,
each ionized \ion{H}{2} region is a spherical bubble of 
radius $R$ with full ionization inside, and neutral hydrogen outside.
We generalize
the parameterization of \citet{wang_hu06} to describe the power spectrum of \ion{H}{2} regions rather than \ion{H}{1} regions and allow
 the \ion{H}{2} bubbles to have a distribution of radii, $P(R)$, that can evolve with redshift.  Note that
 we suppress the redshift arguments to functions such as $P(R)$ where no confusion will arise.
The physical mechanisms of reionization, including what the ionizing sources are and 
how efficiently they can ionize neutral gas surrounding them, are left 
unspecified.  Therefore not all of the parameter space available corresponds
to currently favored, or even physically plausible, models.

In Section~\ref{modelmean} we describe the basic features of the model and the relationship between
the bubble distribution and mean ionization.
In Section~\ref{modelpower} we show how the ionization history and bubble radius distribution 
determines the power spectrum of the ionized hydrogen.   Finally in Section~\ref{avwinln} we
explore the impact of the width of the distribution on the power spectrum.

\subsection{Mean Ionization}
\label{modelmean}

We assume that \ion{H}{2} bubbles populate a large volume $V_{0}$ as a Poisson process with 
fluctuating mean \citep{valageas_etal01,furlanetto_etal04,wang_hu06}.  In addition we allow the bubbles to have an arbitrary distribution of radii $P(R)$.  The volume of each bubble
is $V_b(R)=4\pi R^{3}/3$. Then the ionization fraction is
\begin{equation}
\langle x({\bf r})\rangle_{\rm P}=1-\exp\left[-\int dR\frac{dn_b}{dR}V_b(R)\right],
\label{xpoisson}
\end{equation}
where $dn_b/dR=n_b({\bf r})P(R)$ and $n_b({\bf r})$ is the comoving bubble
number density, and the subscript P denotes averaging over the 
Poisson process.

The average ionization fraction in the volume $V_{0}$ is
\begin{equation}
x_{e}=1-\frac{1}{V_{0}}\int_{V_{0}}d^{3}r\exp\left[-\int dR\frac{dn_b}{dR}V_b(R)\right].\label{avgx}
\end{equation}
We assume the number density of bubbles fluctuates as a biased tracer of the
large scale structure with 
bubble bias $b(R)$:
\begin{equation}
\frac{dn_b}{dR}=\frac{d\bar{n}_b}{dR}\left[1+b(R)\delta_{W}({\bf r})\right],
\label{nbbias}
\end{equation}
where $\delta_{W}({\bf r})=\int d^{3}r'\delta({\bf r'})W_{R}({\bf r}-{\bf r'})$
is the overdensity $\delta$ smoothed with a top-hat window function
\begin{equation}
W_R(k)=\frac{3}{(kR)^3}[\sin(kR)-kR\cos(kR)].
\label{tophat}
\end{equation}  
In our model, we assume that the bubble bias is independent of $R$, $b(R)=b$ for 
simplicity.

Using equation~(\ref{nbbias}) and expanding the exponential
in equation~(\ref{avgx}), we find that the average number density, bubble
volume and ionization fraction are related by
\begin{eqnarray}
 1-x_{e} &\approx& \frac{e^{-\bar{n}_b\langle V_b \rangle}}{V_{0}} \int_{V_{0}}d^{3}r
 \left[1-b\int dR\frac{d\bar{n}_b}{dR}V_b(R)\delta_{W}({\bf r})\right]\nonumber\\
&=& e^{-\bar{n}_b\langle V_b \rangle},
\label{nbvbxe}
\end{eqnarray}
where we define 
\begin{equation}
\langle V_b \rangle\equiv\int dR~P(R)V_b(R),
\end{equation} 
and assume $\int_{V_0} d^{3}r \delta_W({\bf r})=0$ for a sufficiently large volume.  Therefore
any two of \{$x_e, \bar n_b, \langle V_b \rangle$\} specifies the mean ionization model.

\subsection{Ionized Hydrogen Power Spectrum}
\label{modelpower}

The ionization field can be written as 
$x({\bf r})=x_e+\delta x({\bf r})$, where $x_e$ is the spatially 
averaged ionization fraction of equation~(\ref{nbvbxe}).
The ionized hydrogen perturbations have 
contributions from both ionization fluctuations $\delta x$ and 
hydrogen overdensities $\delta$ \citep{hu00}
\begin{equation}
\dhii({\bf r})=\delta x({\bf r})+[x_e+\delta x({\bf r})]\delta({\bf r}).
\label{deltahii}
\end{equation}

The 2-point correlation function of ionized hydrogen, 
$\xi_{\dhii\dhii}(r)=\langle \dhii({\bf r}_1)\dhii({\bf r}_2) \rangle$ 
where $r=|{\bf r}_1-{\bf r}_2|$, is \citep{furlanetto_etal04}
\begin{eqnarray}
\xi_{\dhii\dhii}(r)&=&\xi_{\delta x\delta x}(r)+x_e^2\xi_{\delta\delta}(r)
\label{2ptcorr}\\
&&+\xi_{\delta x\delta~\delta x\delta}(r)+2x_e\xi_{\delta x\delta}(r),
\nonumber
\end{eqnarray}
neglecting terms that are higher order in $\delta$ and $\delta x$ (except 
$\xi_{\delta x\delta~\delta x\delta}$, which turns out to be important 
to this order in the perturbations). 
The \ion{H}{2} power spectrum is the Fourier transform of the 
correlation function,
\begin{equation}
\phii(k)=\int d^3r~e^{i{\bf k}\cdot{\bf r}}\xi_{\dhii\dhii}(r).
\end{equation}
Following \citet{wang_hu06},
we compute the correlation functions in equation~(\ref{2ptcorr}) by 
separating the contributions from one-bubble and two-bubble correlations, 
analogous to the halo model~\citep[see][for a review]{cooray_sheth02}.

\subsubsection{Two-bubble terms}

For the two-bubble contribution to the two-point \ion{H}{2} correlation function, 
the term $\xi^{2b}_{\delta x\delta~\delta x\delta}$ is higher 
order than the other terms and can be dropped~\citep{furlanetto_etal04}. 
We choose to include the density correlation term $x_e^2 \xi_{\delta\delta}$ with the two-bubble 
terms due to their joint dependence on the matter power spectrum $P(k)$. 
The remaining two-bubble terms in the \ion{H}{2} power spectrum 
 are $P^{2b}_{\delta x\delta x}(k)$ and $P^{2b}_{\delta x\delta}(k)$. 
To find these terms, we need an expression for $\delta x({\bf r})$ that 
accounts for the clustering of \ion{H}{2} regions through the bubble bias $b$.

Expanding the exponential in the Poisson-averaged ionization fraction 
equation~(\ref{xpoisson}) gives
\begin{eqnarray}
\langle x({\bf r})\rangle_{\rm P}=1&-&e^{-\bar{n}_b\langle V_b \rangle}\left[1-b\int dR~\bar{n}_bP(R)V_b(R)\delta_{W}({\bf r})\right]\nonumber\\
 = x_{e}&+&(1-x_{e})\bar{n}_bb\int dR~P(R)V_b(R)\delta_{W}({\bf r}),
\end{eqnarray}
so the ionization fraction perturbation is
\begin{eqnarray}
&&\langle\delta x({\bf r})\rangle_{\rm P}=\langle x({\bf r})\rangle_{\rm P}-x_{e}\\
&=&-\frac{(1-x_{e})\ln(1-x_{e})}
{\langle V_b \rangle}b\int dR~P(R)V_b(R)\delta_{W}({\bf r}).\nonumber
\end{eqnarray}

Using this expression for $\delta x$, we can find the correlations 
of $\delta x$ with $\delta x$, and $\delta x$ with $\delta$. The Fourier 
transforms of these correlation functions are
\begin{eqnarray}
P^{2b}_{\delta x\delta x}(k)=\left[(1-x_e)\ln(1-x_e)b\wrktwo\right]^2P(k)\label{p2b1},\\
P^{2b}_{\delta x\delta}(k)=-(1-x_e)\ln(1-x_e)b\wrktwo P(k),\label{p2b2}
\end{eqnarray}
where $P(k)=P_{\delta\delta}(k)$ is the matter power spectrum, and 
the two-bubble window function averaged over the bubble radius 
distribution is defined as
\begin{equation}
\wrktwo=\frac{1}{\langle V_b\rangle}\int dR~V_b(R)P(R)W_R(k),\label{w2b}
\end{equation}
where $W_R(k)$ is the Fourier transform of the real-space top-hat window of
equation~(\ref{tophat}).

The total two-bubble contribution to the power spectrum comes from 
combining the terms in equations~(\ref{p2b1}) and~(\ref{p2b2}) with the 
density fluctuation term $P(k)$, with coefficients as given by 
equation~(\ref{2ptcorr}):
\begin{equation}
\phii^{2b}(k)=\left[(1-x_e)\ln(1-x_e)b\wrktwo-x_e\right]^2P(k).
\label{p2b}
\end{equation}

Finally to complete the description we need a model for the bubble bias $b$.
We assume that the bubbles
are centered on dark matter halos so that the bias of the central halo determines the
bubble bias.   Specifically, we match the desired number density of
\ion{H}{2} bubbles to the number density of halos above a mass threshold $M_{\rm th}$
\begin{equation}
\bar{n}_b=\int_{M_{{\rm th}}}^{\infty}\frac{dM}{M}\frac{dn_h}{d\ln M},
\end{equation}
where 
$dn_h/d\ln M$ is the halo mass function.  In a physical model
$M_{{\rm th}}$ might be associated with the threshold mass for collapse
determined by cooling via atomic hydrogen. 
We use the \citet{sheth_tormen99} mass function for $dn_h/d\ln M$.
The bubbles are then correlated according to the matter power spectrum but with
a bias, 
\begin{equation}
b=\frac{1}{\bar{n}_b}\int_{M_{{\rm th}}}^{\infty}\frac{dM}{M}b_h(M)\frac{dn_h}{d\ln M},\label{bias}
\end{equation}
where the bias of dark matter halos $b_h$ is given by the 
\citet{sheth_tormen99} form.  Note that this association with dark matter
halos is only necessary for modelling the two bubble correlations.  For much
of the parameter space of interest, the one halo term dominates and its contributions
are independent of this association.

\subsubsection{One-bubble terms}

In the one-bubble regime, we can neglect $\xi^{1b}_{\delta x\delta}$ in 
equation~(\ref{2ptcorr}) because we assume complete ionization inside 
bubbles and complete neutrality outside~\citep{wang_hu06}. The 
$\xi_{\delta\delta}$ term in $\xi_{\dhii\dhii}$ is already included 
in the two-bubble contribution to the power spectrum, so only the first 
and third terms in equation~(\ref{2ptcorr}) remain.
For our purposes, the main contribution from 
$\xi^{1b}_{\delta x\delta~\delta x\delta}$ can be approximated by 
$\xi^{1b}_{\delta x\delta x}\xi_{\delta\delta}$~\citep{furlanetto_etal04,wang_hu06}.

The two-point correlation of the total ionization fraction in the 
one-bubble regime can be written \citep{gruzinov_hu98}
\begin{eqnarray}
\langle x({\bf r}_{1})x({\bf r}_{2})\rangle&=&x_e^2+\xi^{1b}_{\delta x\delta x}(r)\label{xx1b}\\
&=&x_{e}P_{2|1}(r)+x_{e}^{2}(1-P_{2|1}(r)),\nonumber
\end{eqnarray}
where $P_{2|1}(r)$ is the probability of the point ${\bf r}_{2}$ being
ionized (in the same bubble as ${\bf r}_{1}$) 
given that ${\bf r}_{1}$ is ionized, and $r=|{\bf r}_1-{\bf r}_2|$. 
The first term is the
probability that both points are ionized in the same bubble, and the
second is the probability that the points are ionized in 
separate bubbles. Then
\begin{equation}
P_{2|1}(r)=\frac{1}{\langle V_b \rangle}\int dRP(R)V_b(R)f\left(\frac{r}{R}\right),
\label{p21}
\end{equation}
where $f(r/R)$ is the probability
of ionization at ${\bf r}_{2}$ given ${\bf r}_{1}$ in an ionization
bubble of radius $R$, with $r=|{\bf r}_{1}-{\bf r}_{2}|$. 

Suppose the point ${\bf r}_1$ is in an ionized bubble of radius $R$ at a 
distance $r_0<R$ from the center, and ${\bf r}_2$ is separated from 
${\bf r}_1$ by $r$.  If we consider two spheres with radii $R$ and $r$, 
with their centers a distance $r_0$ apart, then we can define 
$A(R,r,r_0)$ as the area 
on the sphere with radius $r$ that is in the interior of the sphere with 
radius $R$. The probability that a random point ${\bf r}_1$ inside a bubble 
of radius $R$ is a distance $r_0$ from the center is 
$P(r_0)dr_0=(4\pi r_0^2dr_0)/V_b(R)$. The function $f(r/R)$ is the 
integral over $r_0$ from 0 to $R$ of the fractional area 
$A(R,r,r_0)/(4\pi r^2)$ times $P(r_0)$, which gives
\begin{equation}
f\left(\frac{r}{R}\right)=1-\frac{3}{4}\frac{r}{R}+\frac{1}{16}
\left(\frac{r}{R}\right)^3,~r\leq 2R,
\end{equation}
and $f(r/R)=0$ for $r>2R$.
Assuming there is not significant overlap between bubbles, 
this function $f(r/R)$ is the convolution 
of $V_bW_R(r)$ with itself and its Fourier transform 
is $V_bW_R^2(k)$~\citep{wang_hu06}:
\begin{eqnarray}
f\left(\frac{r}{R}\right)&=&\frac{1}{V_b(R)}\int d^{3}r'V_b^{2}(R)W_{R}({\bf r}-{\bf r'})W_{R}({\bf r'})\nonumber\\
&=&V_b(R)\int\frac{d^{3}k}{(2\pi)^{3}}W_{R}^{2}(k)e^{-i{\bf k}\cdot{\bf r}}.
\end{eqnarray}
The overlap between bubbles will be small if 
$\bar{n}_b\langle V_b\rangle\ll 1$, which corresponds to 
$x_e\ll 0.63$ by equation~(\ref{nbvbxe}). This assumption should be 
valid until near the end of reionization.

Using this expression for $f(r/R)$ with equations~(\ref{xx1b}) 
and~(\ref{p21}), the ionization correlation function is
\begin{eqnarray}
\xi^{1b}_{\delta x\delta x}(r)&=&(x_{e}-x_{e}^{2})P_{2|1}(r),\label{xi1bdxdx}\\
P_{2|1}(r) &=&\frac{1}{\langle V_b \rangle}\int dR~P(R)V_b^{2}(R) \, \int\frac{d^{3}k}{(2\pi)^{3}}W_{R}^{2}(k)e^{-i{\bf k}\cdot{\bf r}}.\nonumber
\end{eqnarray}
We can take the Fourier transforms of 
$\xi^{1b}_{\delta x\delta x}$ and 
$\xi^{1b}_{\delta x\delta x}\xi_{\delta\delta}$ to get
 the 1-bubble terms in the power spectrum
\begin{eqnarray}
P^{1b}_{\delta x\delta x}(k)&=&\frac{x_{e}-x_{e}^{2}}{\langle V_b \rangle}\int dR~V_b^{2}(R)P(R)W_{R}^{2}(k),\label{p1b1}\\
P^{1b}_{\delta x\delta~\delta x\delta}(k)&=&(x_{e}-x_{e}^{2})\tilde{P}(k).\label{p1b2}
\end{eqnarray}

The first term, equation~(\ref{p1b1}), is associated with the shot noise 
of the bubbles. In the second term, equation~(\ref{p1b2}), 
$\tilde{P}(k)$ is the convolution of the matter power spectrum $P(k)$ with 
the square of the one-bubble averaged window function,
\begin{equation}
\tilde{P}(k)=\langle V_b\rangle\int\frac{d^3k_1}{(2\pi)^3}
\wrkone P(|{\bf k}-{\bf k}_1|),
\end{equation}
where the one-bubble window function averaged over the 
bubble radius distribution is defined by
\begin{equation}
\wrkone=\frac{1}{\langle V_b\rangle^2}\int dR~V_b^2(R)P(R)W^2_R(k).\label{w1b}
\end{equation}
As in \citet{wang_hu06}, we simplify our calculations by approximating 
$\tilde{P}(k)$ by interpolating between its small and large $k$ limits:
\begin{equation}
\tilde{P}(k)\approx \frac{P(k)\langle V_b\rangle\langle \sigma_R^2\rangle}
{[(P(k))^2+(\langle V_b\rangle\langle \sigma_R^2\rangle)^2]^{1/2}},
\end{equation}
where $\langle \sigma_R^2\rangle$ is the top-hat smoothed density variance 
averaged over the bubble radius distribution
\begin{equation}
\langle \sigma_R^2\rangle=\frac{1}{\langle V_b\rangle^2}\int dR~V_b^2(R)
P(R)\sigma_R^2.
\end{equation}

The total one-bubble contribution to the \ion{H}{2} power spectrum is the sum 
of equations~(\ref{p1b1}) and~(\ref{p1b2}),
\begin{equation}
\phii^{1b}(k)=x_e(1-x_e)\left[\langle V_b\rangle \wrkone+\tilde{P}(k)\right].
\label{p1b}
\end{equation}

\subsection{Bubble Radius Distribution}
\label{avwinln}

\begin{figure}
\epsscale{1.0}
\centerline{\psfig{file=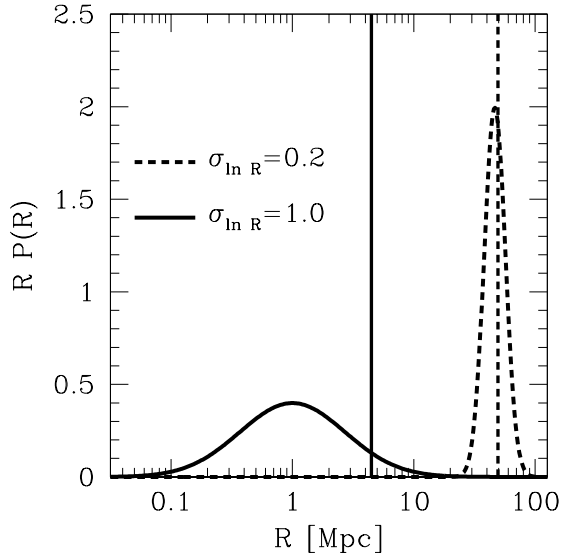, width=3.0in}}
\caption{\ion{H}{2} bubble radius distributions for $\slnr=0.2$ (\emph{dashed}) and 
$\slnr=1.0$ (\emph{solid}) with the same effective radius, $R_{\rm eff}=55$~Mpc. The vertical lines show the location of the volume 
weighted radius $R_0$ for each distribution.
\vskip 0.25cm}
\label{raddist}
\end{figure}

To evaluate the power spectrum, we need to specify the bubble radius distribution which determines
the averaged window functions $\wrktwo$ and $\wrkone$.
Motivated by the analytic model of \citet{furlanetto_etal04} and 
numerical simulations \citep{zahn_etal06}, we assume here a 
log normal distribution with width $\slnr$:
\begin{equation}
P(R)=\frac{1}{R}\frac{1}{\sqrt{2\pi\slnr^{2}}}e^{-[\ln(R/\bar{R})]^{2}/2\slnr^{2}}.
\end{equation}
This distribution is shown in Figure~\ref{raddist} for  
$\slnr=0.2$ and $\slnr=1.0$. 

\begin{figure}
\epsscale{1.0}
\centerline{\psfig{file=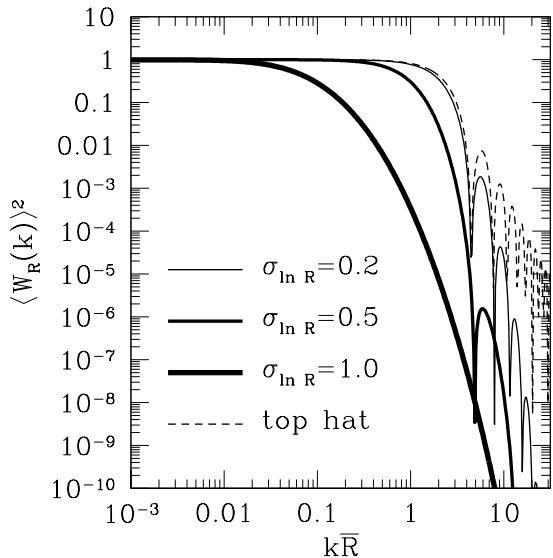, width=3.0in}}
\caption{Squared two-bubble averaged window functions for log normal 
bubble radius distributions with widths $\slnr=0.2$ (\emph{thin}), 0.5 
(\emph{medium}), and 1.0 (\emph{thick}). The square of a top-hat 
window function is plotted for comparison (\emph{dashed}).
\vskip 0.25cm}
\label{win2b}
\end{figure}

\begin{figure}
\epsscale{1.0}
\centerline{\psfig{file=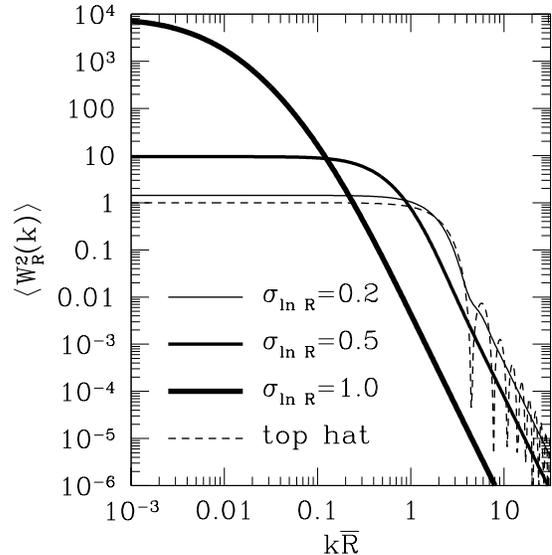, width=3.0in}}
\caption{Same as Figure~\ref{win2b}, but for one-bubble averaged window 
functions.
\vskip 0.5cm}
\label{win1b}
\end{figure}

The one and two bubble terms in the power spectrum possess different weights over this distribution.
From equation~(\ref{w2b}) using a log normal $P(R)$, the two-bubble 
averaged window function is
\begin{eqnarray}
\wrktwo&=&\frac{3e^{-9\slnr^2/2}}{\sqrt{2\pi}\slnr}\int_{0}^{\infty}
d\chi~\chi^2e^{-\ln^{2}\chi/2\slnr^{2}}\label{w2bln}\\
&&\times\frac{\sin\kappa\chi-\kappa\chi\cos\kappa\chi}{(\kappa\chi)^3}\,\nonumber
\end{eqnarray}
where $\kappa\equiv k\bar{R}$ and $\chi\equiv R/\bar{R}$.  
Similarly, for the one-bubble terms, equation~(\ref{w1b}) becomes
\begin{eqnarray}
\wrkone&=&\frac{9e^{-9\slnr^2}}{\sqrt{2\pi}\slnr}\int_{0}^{\infty}
d\chi~\chi^5e^{-\ln^{2}\chi/2\slnr^{2}}\label{w1bln}\\
&&\times\frac{(\sin\kappa\chi-\kappa\chi\cos\kappa\chi)^2}{(\kappa\chi)^6}.\nonumber
\end{eqnarray}
For a given value of $\slnr$, each of these averaged window functions 
depends only on $\kappa=k\bar{R}$ and not on $k$ or $\bar{R}$ separately.
Figures~\ref{win2b} and~\ref{win1b} show the window functions for various 
choices of $\slnr$, evaluated by numerically integrating the expressions 
in equations~(\ref{w2bln}) and~(\ref{w1bln}).

It is useful
to examine the limiting behavior of these window functions to determine the characteristic scale
in the average.
Using the fact that $\lim_{k\rightarrow 0}W_R(k)=1$, we see from 
equation~(\ref{w2b}) that the two-bubble window function $\wrtwo$ 
also approaches unity in the limit of small $k$.   In the oscillatory high $k$ regime, 
the integration will sharply
suppress contributions.  Thus the weight in the distribution
simply reflects the average bubble volume and it is useful to define a volume weighted radius $R_0$
\begin{equation}
\langle V_b\rangle\equiv\frac{4\pi}{3}R_0^3,\label{R0def}
\end{equation}
which is larger than $\bar{R}$ (see Figure~\ref{raddist})
\begin{equation}
R_0=e^{3\slnr^2/2}\bar{R}.
\end{equation}

The $k\rightarrow 0$ limit of equation~(\ref{w1b}) for the
one-bubble window function is
\begin{equation}
\lim_{k\rightarrow 0}\wrkone=\frac{\langle V_b^2\rangle}
{\langle V_b\rangle^2}=e^{9\slnr^2},\label{w1bsmallk}
\end{equation}
where the second equality holds for the case where $P(R)$ is the 
log normal distribution.
 For large $k$, 
$W_R^2(k)\approx 9(kR)^{-4}\cos^2(kR)$. 
Since $\cos^2(kR)$ 
oscillates rapidly in this limit compared with the variation of 
$V_b^2P(R)R^{-4}$, we can 
replace $(\sin\kappa\chi-\kappa\chi\cos\kappa\chi)^2$
by $1/2$ in the integral in equation~(\ref{w1bln}). Then we have
\begin{eqnarray}
\lim_{k\rightarrow\infty}\wrkone&\approx&
\frac{9e^{-9\slnr^2}}{2\sqrt{2\pi}\slnr\kappa^4}\int
d\chi~\chi e^{-\ln^{2}\chi/2\slnr^{2}}\label{w1blargek}\nonumber\\
&=&\frac{9}{2}e^{-7\slnr^2}(k\bar{R})^{-4}.
\end{eqnarray}

To estimate the characteristic scale in the one-bubble terms, we 
take the 
value of $k$ where $\wrone$ transitions between these two limits. 
This wavenumber, $k_{\rm eff}$, can be estimated by requiring that 
\begin{equation}
\left.\lim_{k\rightarrow 0}\wrkone\right|_{k=k_{\rm eff}}=
\left.\lim_{k\rightarrow\infty}\wrkone\right|_{k=k_{\rm eff}}.
\end{equation}
Using equations~(\ref{w1bsmallk}) and~(\ref{w1blargek}), we find
\begin{equation}
k_{\rm eff}=\left(\frac{9}{2}\right)^{1/4}R_0^{-1}e^{-2.5\slnr^2}.
\end{equation}
Therefore, we expect the main contributions to arise from
scales
\begin{equation}
R_{{\rm eff}}\equiv  \bar{R}  e^{4\slnr^2} = R_0 e^{2.5\slnr^2} .
\label{Reffdef}
\end{equation}
Note that the two distributions in Figure~\ref{raddist} have the same $R_{\rm eff}$
even though their $\bar R$ differ by a factor of $\sim 50$. 
In the $\slnr\rightarrow 0$ limit, $R_{\rm eff}=\bar{R}=R_0$ so all 
the polarization power comes from bubbles with radii equal to $R_{\rm eff}$. 
For larger values of $\slnr$, we find that most of the power is still 
due to bubbles with $R\sim R_{\rm eff}$.

\section{Parameterization of Reionization Models}
\label{parameterization}

The framework for inhomogeneous reionization 
described in the previous section requires the specification of three functions of
redshift: the ionization fraction
$x_e(z)$, the volume weighted bubble radius $R_0(z)$  
defined by 
equation~(\ref{R0def}), and the width of the log normal radius distribution
$\slnr(z)$.   In this section we discuss several parameterizations of these functions that
highlight their impact on the \emph{B}-modes.

We begin by discussing the leading order shot noise term contributing to the \emph{B}-mode spectrum
and how it determines the shape of the spectrum in Section~\ref{Bspectrum}.
We parameterize the range of redshifts for which $0 < x_e < 1$ by $z_i$ and $z_f$, the
redshift at the beginning and end of reionization respectively. In Section~\ref{xesection}, we explore the impact of various choices for $x_e(z)$ during this
epoch, including histories that maximize the $B$-modes and integral constraints from the total 
optical depth $\tau_*$.
  Finally we take two
representative forms for $R_0(z)$ in Section~\ref{parrsig}, a constant radius set by $R_{\rm eff}$ the characteristic
radius of the bubbles for the dominant shot noise term, and a constant number density
of bubbles $\bar{n}_b$ which grow in radius.   For simplicity we take a constant $\slnr$ throughout.

\subsection{Approximate B-mode Spectrum}
\label{Bspectrum}

\begin{figure}
\epsscale{1.0}
\centerline{\psfig{file=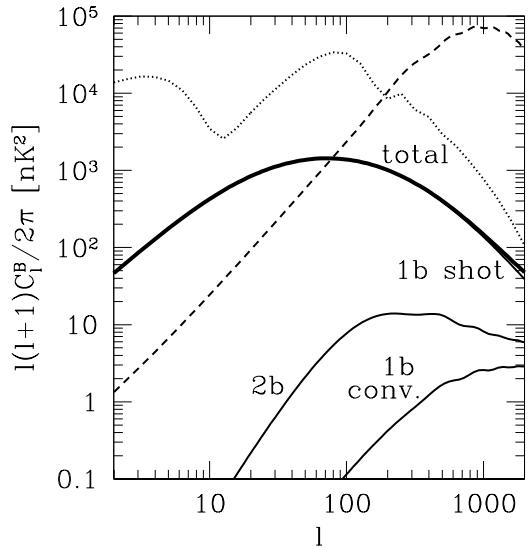, width=3.0in}}
\caption{An example of a $B$-mode polarization spectrum from 
inhomogeneous reionization (\emph{solid, thick}), and the contributions to the 
spectrum from each component of the \ion{H}{2} power spectrum in 
equation~(\ref{phiicomps}) (\emph{thin, solid curves} as labelled): the one-bubble shot noise term, 
the one-bubble convolution term, and the two-bubble 
term. The reionization model that produces this 
polarization (model 19 in Table~\ref{tabrsig}) has a linear ionization history, constant $\bar{n}_b$, $z_f=6$, 
$z_i=18.4$, $\tau_*=0.12$, $\slnr=1.0$, and $R_0(z_f)=50$ Mpc ($R_{\rm eff}(z_f)=609$ Mpc). 
The lensing $B$-modes for the assumed cosmology (\emph{dashed}) and 
the gravitational wave $B$-modes with tensor-to-scalar ratio 
$r=0.5$ (\emph{dotted}) are plotted for comparison.
\vskip 1.0cm}
\label{clbcomps}
\end{figure}

Different components of the \ion{H}{2} power spectrum can vary greatly in how 
much they contribute to $\cclb$ at a particular scale. We study the 
relative contributions by separating the power spectrum into the 
one-bubble shot noise term, the one-bubble convolution term, and the 
two-bubble term
\begin{equation}
\phii=\pshot+\pconv+P^{2b}_{\dhii} \label{phiicomps}
\end{equation}
and integrating equation~(\ref{clrei}) replacing $\phii$ by each of these 
three terms.  Figure~\ref{clbcomps} shows these contributions to the 
total $B$-mode spectrum for a particular reionization model.
The lensing and gravitational wave $B$-modes plotted in Figure~\ref{clbcomps} 
show that $B$-modes from inhomogeneous reionization would be an important
contaminant for gravitational waves
 if they have a large amplitude at $\ell\lesssim 100$, so 
we sometimes focus on the $B$-mode amplitude at $\ell\sim 100$.

For all values of model parameters we consider, the contribution to 
$\cclb$ from the convolution term, $\pconv$, is less than the 
contributions from the other two terms.

For some choices of parameters, 
the contribution from $P^{2b}_{\dhii}$ is dominant over a range of $\ell$ 
values. In these cases, however, the $B$-mode amplitude is never 
larger than a few tens of nK$^2$. The amplitude of the two-bubble 
contribution is limited by the fact that $\wrkone$ is never greater than 
1, and because $P(k)$ decreases with redshift.

All of the models with large $B$-modes ($\cclb \gtrsim 100$ nK$^2$) 
at 
$\ell\sim 100$ are dominated by the one-bubble shot noise term in the 
\ion{H}{2} power spectrum, $\pshot$ (e.g., the model in Figure~\ref{clbcomps}). Note that this is the only term in the \ion{H}{2} power spectrum that is 
independent of the matter power spectrum.

Since the shot noise term dominates the polarization spectrum, we can 
understand much of the dependence on reionization parameters by 
approximating the $B$-mode spectrum by its shot noise contribution. 
For these models $\cclb$ is well described as a 
single, wide peak. We will often give our results by stating the 
multipole of this peak, $\lpeak$, and its amplitude $(\cclb)_{\rm peak}$.

The scale $\lpeak$ is primarily determined 
by the bend in the averaged one-bubble window function $\wrone$, which 
is set by $R_{\rm eff}$ as described in Section~\ref{avwinln}.  More 
specifically $\lpeak = k_{\rm eff} / D_A$ and since the 
angular distance varies slowly during reionization, the shot 
noise-dominated $B$-mode power should peak
at 
\begin{equation}
\lpeak\propto R_{\rm eff}^{-1} \label{lpkrsig}
\end{equation}
with power law behavior as $\cclb\sim\ell^2$ on 
large scales and $\cclb\sim\ell^{-2}$ on small scales. 
Because 
of the crucial role $R_{\rm eff}$ plays in setting the multipole at 
which the polarization spectrum peaks, we present our results in 
terms of $R_{\rm eff}$ rather than $R_0$.    

The amplitude of the peak depends on multiple parameters of the ionization model
as we describe in Section~\ref{pardep}.

\subsection{Ionization Fraction}\label{xesection}

To explore the dependence of the $B$-modes on $x_e(z)$, we use a variety 
of types of ionization histories.

We assume that $x_e=0$ for $z>z_i$ and $x_e=1$ for $z<z_f$. During the epoch 
of reionization, we require $0\leq x_e\leq 1$.
The optical depth places an additional constraint on $x_e(z)$. The optical 
depth out to redshift $z$ is
\begin{equation}
\tau(z)=\sigma_Tn_{p,0}\int_0^zdz'\frac{(1+z')^2}{H(z')}x_e(z').
\label{tauz}
\end{equation}
We can use the range of observed values for the total optical
depth $\tau_*=\tau(z\geq z_i)$ 
to constrain $x_e(z)$ through equation~(\ref{tauz}).
Note that the epoch of reionization can start 
no later than $z_i\approx 12.4$ if $\tau_*=0.12$, or $z_i\approx 7.7$ 
if $\tau_*=0.06$.

The polarization signal we are concerned with only comes from redshifts $z_f<z<z_i$, so 
we will define the optical depth from the epoch of reionization
$\tau_{\rm rei}\equiv \tau_*-\tau(z_f)$, where $\tau(z_f)$ depends only 
on $z_f$ and cosmological parameters that we hold fixed.

To derive approximate expressions in the following sections, we will often 
assume that the universe is matter dominated so that 
$H(z)\approx H_0\sqrt{\Omega_{\rm m}}(1+z)^{3/2}$. This is a good approximation 
during the epoch of reionization.

\subsubsection{Constant Ionization Fraction}

The simplest model is a constant value of 
$x_e$ throughout the epoch of reionization. For a fixed value of $\tau_*$, 
the ionization fraction in this model is
\begin{equation}
x_e=\frac{3\tau_{\rm rei}}{2\tau_0}\left[(1+z_i)^{3/2}-(1+z_f)^{3/2}\right]^{-1}\,,
\label{constxe}
\end{equation}
where $\tau_0=\sigma_Tn_{p,0}/(\sqrt{\Omega_{\rm m}}H_0)$. 

Although a model with constant $x_e$ is far from being a realistic 
model of reionization, it is a useful simplifying assumption for 
finding the approximate dependence of $\cclb$ on other parameters of the 
model.

\subsubsection{Linear Ionization Fraction}

\begin{figure}
\epsscale{1.0}
\centerline{\psfig{file=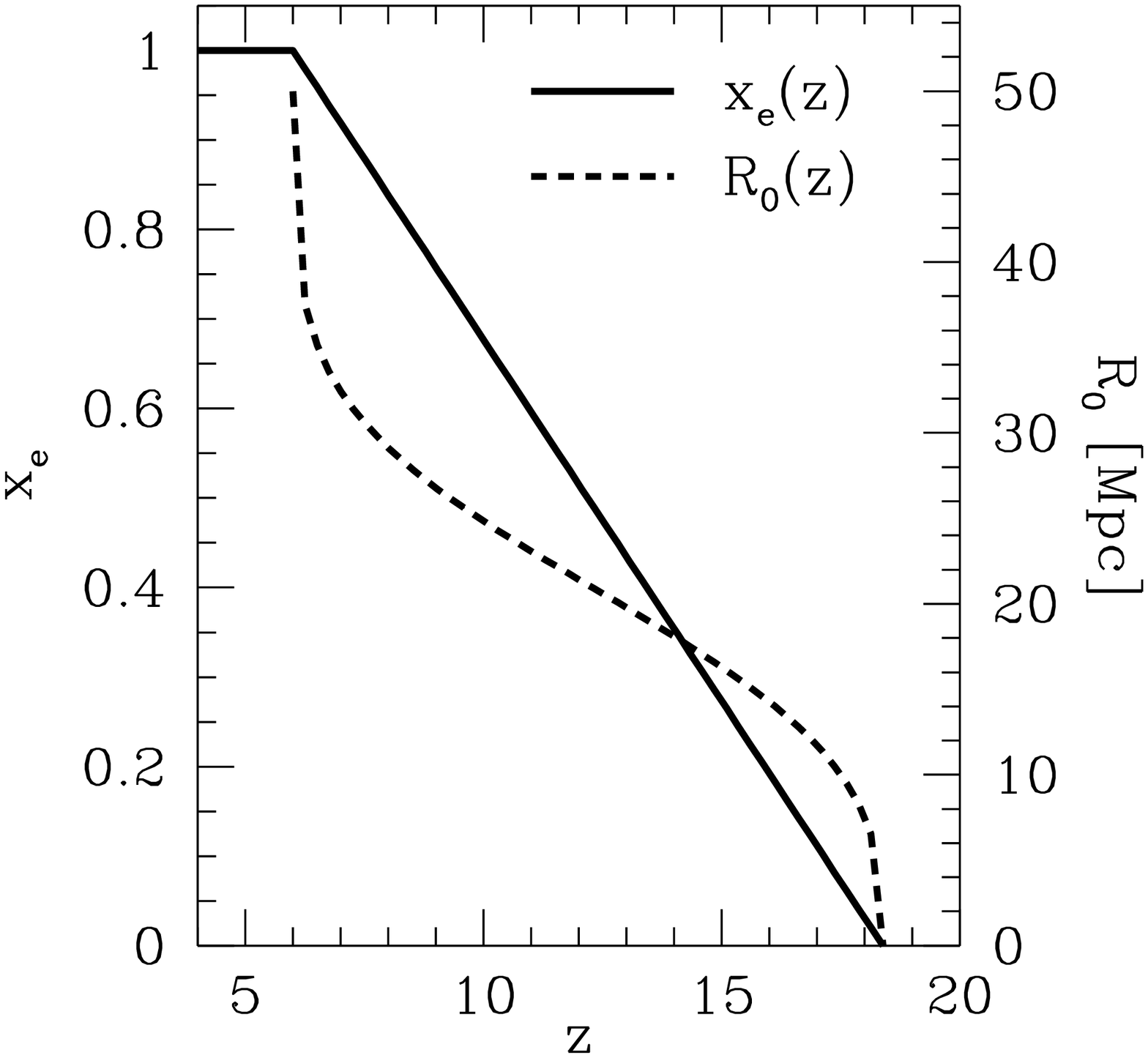, width=3.0in}}
\centerline{\psfig{file=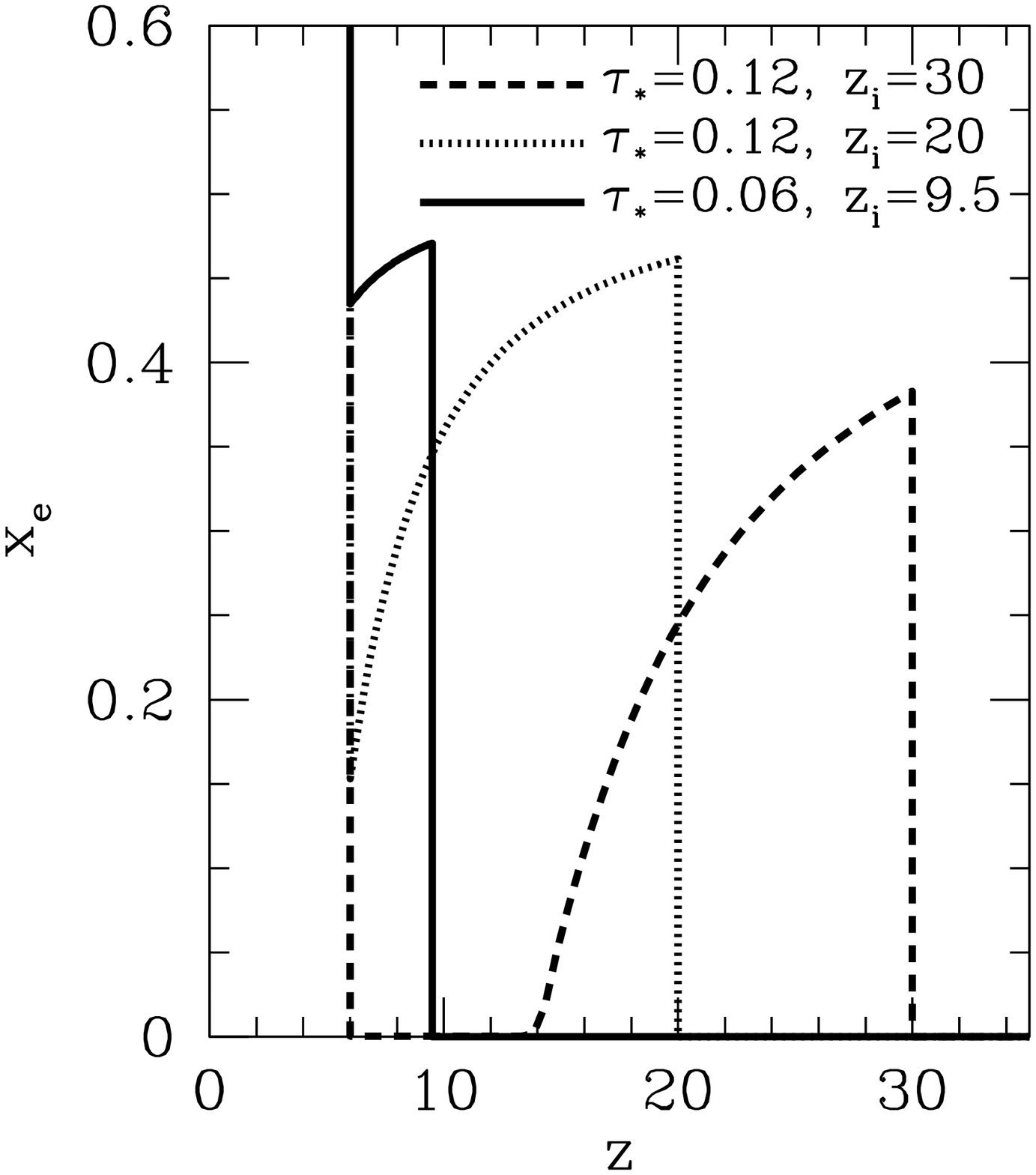, width=3.0in}}
\caption{\emph{Upper panel}: Ionization history and characteristic bubble radius 
for a linear $x_e(z)$ model with $z_f=6$, $z_i=18.4$, $\tau_*=0.12$, and fixed $\bar{n}_b$ 
(model 19 in Table~\ref{tabrsig}). \emph{Lower panel}: Ionization histories for 
three maximal-$x_e(z)$ models (models 8-10 in Table~\ref{tabxe}). The 
ionization fraction in all three models increases to 1 for $z\leq 6$.
\vskip 0.5cm}
\label{xezfig}
\end{figure}

As a more realistic model of reionization  
we consider an ionization fraction that decreases linearly with redshift,
\begin{equation}
x_e(z)=1-\frac{z-z_f}{z_i-z_f}.
\end{equation}
For a fixed value of $\tau_*$, the redshifts $z_i$ and $z_f$ cannot be varied independently 
since they are related through equation~(\ref{tauz}). In our linear 
ionization histories, we fix $\tau_*$ and $z_f$ and determine the 
required value for $z_i$.
The upper panel of Figure~\ref{xezfig} shows a linear $x_e(z)$ for a 
model with $\tau_*=0.12$ and $z_f=6$.

\subsubsection{Maximal B-Mode Ionization History}

If we approximate the \ion{H}{2} power spectrum by 
the shot noise term $\pshot$, and use the fact that $D_A(z)$, $Q_{{\rm rms}}(z)$, 
and $\tau(z)$ vary slowly with redshift during reionization to pull these 
factors out of the integral in equation~(\ref{clrei}), then the 
dependence of the $B$-mode amplitude on the ionization history is 
approximately
\begin{equation}
\cclb\sim\int_{z_f}^{z_i}dz(1+z)^{5/2}x_e(z)[1-x_e(z)].\label{clbapprox}
\end{equation}
We can use variational calculus to determine what form of $x_e(z)$ 
maximizes this integral, giving the largest $B$-mode amplitude in this 
approximation.

Fixing the total optical depth $\tau_*$ 
(and therefore $\tau_{\rm rei}=\tau_*-\tau(z_f)$) 
gives an integral constraint on $x_e(z)$,
\begin{equation}
\frac{\tau_{\rm rei}}{\tau_0}=\int_{z_f}^{z_i}dz(1+z)^{1/2}x_e(z),
\label{tauconstraint}
\end{equation}
so we can define
\begin{equation}
f(z,x_e,x_e')=(1+z)^{5/2}x_e(1-x_e)+\lambda (1+z)^{1/2}x_e,
\end{equation}
where $x_e'=dx_e/dz$ and $\lambda$ is a Lagrange multiplier. Since 
this function is independent of $x_e'$, the Euler-Lagrange equation 
is just $\partial f/\partial x_e=0$. Solving for $x_e(z)$ gives
\begin{equation}
x_e(z)=\frac{1}{2}+\frac{\lambda}{2}(1+z)^{-2}.\label{xemax}
\end{equation}

The ionization fraction given by this equation approaches $1/2$ at high redshift. 
For lower 
values of $z$, $x_e$ must stay between by 0 and 1. 
These limits are satisfied for redshifts $z\geq |\lambda|^{1/2}-1$. 
If $z_f$ satisfies this inequality, then $x_e(z)$ as given by 
equation~(\ref{xemax}) is between 0 and 1 during the entire epoch of 
reionization. If not, then for redshifts $z_f<z<|\lambda|^{1/2}-1$, 
$x_e$ is saturated at either 0 or 1 depending on the sign of $\lambda$.

The values we assume for $\tau_*$, $z_i$, and $z_f$ determine the 
sign of $\lambda$. 
If $\lambda>0$, then $x_e>1/2$ for all $z$, so $\tau_{\rm rei}/\tau_0$ must 
be greater than it would be with constant $x_e=1/2$. If $\lambda<0$ the 
ionization fraction is always less than $1/2$ and $\tau_{\rm rei}/\tau_0$ is 
less than the constant $x_e=1/2$ value. We can define an optical depth 
$\tau_{\lambda}$ corresponding to the case $x_e(z)=1/2$
\begin{equation}
\frac{\tau_{\lambda}(z_i,z_f)}{\tau_0}=\frac{1}{3}\left[(1+z_i)^{3/2}
-(1+z_f)^{3/2}\right],\label{taulambda}
\end{equation}
so that $\lambda$ in equation~(\ref{xemax}) must have the same sign as 
$\tau_{\rm rei}-\tau_{\lambda}$. For $z_f=6$ and $z_i=20$, 
$\tau_{\lambda}\approx 0.1$; for $z_f=6$ and $z_i=30$, 
$\tau_{\lambda}\approx 0.2$.

After determining the sign of $\lambda$ for particular choices of 
$\tau_*$, $z_i$, and $z_f$, the magnitude of $\lambda$ is fixed by 
the optical depth constraint, equation~(\ref{tauconstraint}), using 
equation~(\ref{xemax}) for $x_e(z)$ over the appropriate range of 
redshifts. The resulting ionization histories for three combinations 
of $\tau_*$ and $z_i$ are plotted in the lower panel of 
Figure~\ref{xezfig}.

The maximal-polarization ionization histories are even less 
realistic than those described in the previous sections, since they have a 
significant ionization fraction ($x_e\lesssim 1/2$) at high redshift, and 
$x_e$ can be small or even zero near the end of reionization, 
if $\lambda<0$. These models are useful, however, because the 
$B$-mode amplitudes associated with them provide an upper limit 
that $B$-modes from more realistic reionization scenarios must 
lie below.

Note that due to the approximations leading to equation~(\ref{clbapprox}), 
the actual ionization history that maximizes the $B$-mode amplitude 
will differ slightly from the one derived here in equation~(\ref{xemax}). 
The results of Monte Carlo models that vary $x_e(z)$ in each redshift 
bin suggest that the true maximal $x_e(z)$ is not very different from 
the approximation used here. The $B$-mode amplitude of the true 
maximal-$x_e$ model is likely only a few percent larger than the model 
of equation~(\ref{xemax}), and even if the amplitude were much larger it 
would be only for a finely tuned (and unrealistic) ionization history missed in
the Monte Carlo search. 
It is reasonable, therefore, to take the $B$ polarization produced by 
models of the approximate form derived here as an effective upper limit 
on the amplitude that can be reached by varying $x_e(z)$ with $\tau_*$ and 
all other reionization parameters held fixed.

\begin{figure}
\epsscale{1.0}
\centerline{\psfig{file=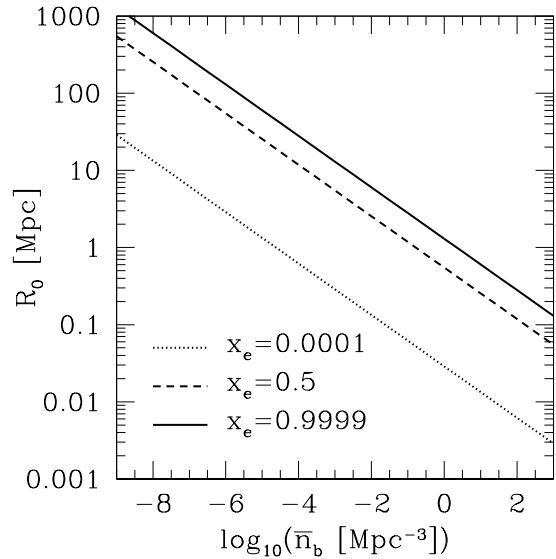, width=3.0in}}
\caption{Volume averaged bubble radius $R_0$ versus bubble comoving 
number density for ionization fractions $x_e=0.0001$, 0.5, and 0.9999. 
Fixed $R_0$ models correspond to horizontal trajectories between 
$x_e\approx 0$ and $x_e\approx 1$, and fixed $\bar{n}_b$ models are 
vertical trajectories in this plot.
\vskip 0.5cm}
\label{rnb}
\end{figure}

\subsection{Bubble Radius}\label{parrsig}

We consider two ways of modelling the evolution of the bubble 
size distribution. 
In one, we assume a fixed comoving bubble number 
density $\bar{n}_b$ throughout reionization, so that the global ionization 
fraction increases only due to bubble growth. The normalization of the 
bubble radius evolution is set by specifying the comoving 
characteristic bubble radius at the end of reionization, $R_{{\rm eff},f}$. 
The evolution of $R_{\rm eff}$ in a fixed-$\bar{n}_b$ model with a linear 
ionization history is shown in the upper panel of Figure~\ref{xezfig}.

The second type of model has a fixed distribution of radii, so 
$R_{\rm eff}$ is independent of $z$. In this model, the ionization fraction 
increases as more ionizing sources collapse and turn on, while the size 
of individual bubbles remains constant. 

In the radius-number density 
plane, these models can be thought of as trajectories along a horizontal 
path from $x_e=0$ to $x_e=1$ for fixed $R_{\rm eff}$, and vertical trajectories for 
fixed $\bar{n}_b$ (Figure~\ref{rnb}). We assume throughout that $\slnr$ does not 
vary with redshift, so the evolution of $R_0$ and $\bar{R}$ is the same as 
that of $R_{\rm eff}$ up to a $\slnr$-dependent normalization.

\begin{center}
\begin{table}
\caption{Multipoles and amplitudes of $B$ polarization spectra peaks 
for reionization models with different types of ionization histories: 
constant (C), linear (L), and for maximal $B$-modes (M).}
\begin{tabular}{|c|ccccccc|cc|}
 & & & & & fixed & & $R_{{\rm eff},f}$ & & $(\cclb)_{{\rm peak}}$ \\
Model & $\tau_*$ & $z_i$ & $z_f$ & $x_e(z)$ & $R_{\rm eff}/\bar{n}_b$ & $\slnr$ & [Mpc] & $\lpeak$ & [nK$^2$] \\
\hline
1 & 0.12 & 20 & 6 & C & both & 1.0 & 609 & 28 & 8836 \\
2 & 0.12 & 30 & 6 & C & both & 1.0 & 609 & 30 & 23050 \\
3 & 0.12 & 40 & 6 & C & both & 1.0 & 609 & 32 & 45049 \\
4 & 0.06 & 9.3 & 6 & L & $R_{\rm eff}$ & 1.0 & 609 & 24 & 422 \\
5 & 0.12 & 18.4 & 6 & L & $R_{\rm eff}$ & 1.0 & 609 & 27 & 4325 \\
6 & 0.12 & 16.6 & 8 & L & $R_{\rm eff}$ & 1.0 & 609 & 27 & 2944 \\
7 & 0.12 & 14.8 & 10 & L & $R_{\rm eff}$ & 1.0 & 609 & 27 & 1600 \\
8 & 0.06 & 9.5 & 6 & M & $R_{\rm eff}$ & 1.0 & 609 & 24 & 692 \\
9 & 0.12 & 20 & 6 & M & $R_{\rm eff}$ & 1.0 & 609 & 28 & 8989 \\
10 & 0.12 & 30 & 6 & M & $R_{\rm eff}$ & 1.0 & 609 & 31 & 27611 \\
11 & 0.12 & 40 & 6 & M & $R_{\rm eff}$ & 1.0 & 609 & 32 & 62667 \\
\end{tabular}
\vskip 0.75cm
\label{tabxe}
\end{table}
\end{center}

\section{$B$-mode Parameter Dependence}\label{pardep}
\label{bmodes}
 
In the previous section, we parameterized the ionization history $x_e(z)$ and the effective
bubble size $R_{\rm eff}(z)$ under several simple schemes that were designed to bring out the
critical properties of a reionization model for the $B$-mode spectrum.  In this section
we explore these properties and provide scaling relations for the impact of reionization 
parameters on $B$-modes.

\subsection{Ionization History}

Table~\ref{tabxe} gives the $B$-mode peak multipoles and amplitudes 
for reionization models with the different ionization histories 
described in Section~\ref{xesection}. For fixed values of $\tau_*$, 
$z_i$, $z_f$, $\slnr$, and $R_{\rm eff}(z)$, we find that changing the shape 
of the ionization history has only a mild effect on the $B$-mode 
spectrum compared with the effects of the other reionization 
parameters.

For example, compare models 1, 5, and 9, which all 
have $\tau_*=0.12$, $z_i\approx 20$, $z_f=6$, $\slnr=1.0$, and 
$R_{\rm eff}=609$ comoving Mpc ($R_0=50$ Mpc). All of these models produce 
$B$-modes that peak at $\lpeak\approx 28$. There is about a factor of 
two difference between the largest peak amplitude of these four 
models (model 9, maximal $x_e(z)$ with $\cclbpeak=8989$ nK$^2$), and 
the smallest (model 5, linear $x_e(z)$ with $\cclbpeak=4325$ nK$^2$).

The results are similar if we adopt different values for the 
reionization parameters other than $x_e(z)$.
For example, models 4 and 8, 
both with $\tau_*=0.06$, $z_i\approx 9.5$, $z_f=6$, and the 
same $R_{\rm eff}(z)$, have the same peak multipole, and their 
$B$-mode amplitudes differ by less than a factor of two between the 
maximal and linear $x_e(z)$ models.

Looking at the entire set of models presented in Table~\ref{tabxe}, there 
is a wide variety of $B$-mode peak scales and amplitudes. However, the 
largest differences occur between models with different choices of $\tau_*$, 
$z_i$, $z_f$, or $R_{\rm eff}(z)$.

\begin{center}
\begin{table}
\caption{Multipoles and amplitudes of $B$ polarization spectra peaks 
for reionization models varying $R_{\rm eff}$ and $\slnr$.}
\begin{tabular}{|c|ccccccc|cc|}
 & & & & & fixed & & $R_{{\rm eff},f}$ & & $(\cclb)_{{\rm peak}}$ \\
Model & $\tau_*$ & $z_i$ & $z_f$ & $x_e(z)$ & $R_{\rm eff}/\bar{n}_b$ & $\slnr$ & [Mpc] & $\lpeak$ & [nK$^2$] \\
\hline
12 & 0.12 & 30 & 6 & C & both & 1.0 & 50 & 400 & 1928 \\
13 & 0.12 & 30 & 6 & C & both & 0.5 & 50 & 407 & 4249 \\
14 & 0.12 & 30 & 6 & C & both & 0.2 & 50 & 450 & 6276 \\
15 & 0.12 & 30 & 6 & C & both & 1.0 & 500 & 37 & 19057 \\
16 & 0.12 & 30 & 6 & C & both & 0.5 & 500 & 40 & 42711 \\
17 & 0.12 & 30 & 6 & C & both & 0.2 & 500 & 45 & 63289 \\
18 & 0.06 & 9.3 & 6 & L & $\bar{n}_b$ & 1.0 & 609 & 59 & 161 \\
19 & 0.12 & 18.4 & 6 & L & $\bar{n}_b$ & 1.0 & 609 & 72 & 1544 \\
20 & 0.12 & 16.6 & 8 & L & $\bar{n}_b$ & 1.0 & 609 & 69 & 1092 \\
21 & 0.12 & 14.8 & 10 & L & $\bar{n}_b$ & 1.0 & 609 & 66 & 619 \\
\end{tabular}
\vskip 0.25cm
\label{tabrsig}
\end{table}
\end{center}

\subsection{\ion{H}{2} Bubble Size Distribution}
\label{sizedist}

\begin{figure}
\epsscale{1.0}
\centerline{\psfig{file=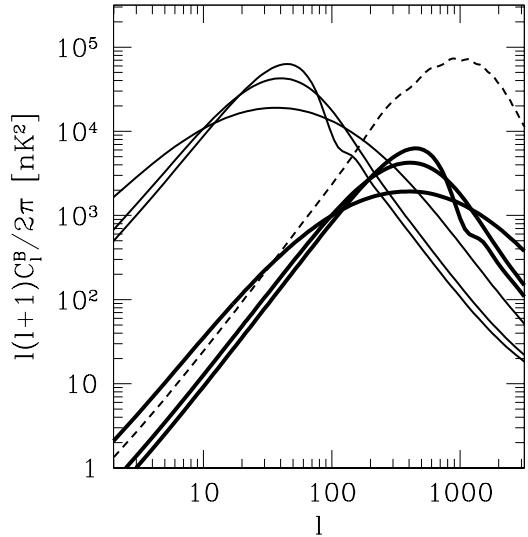, width=3.0in}}
\caption{$B$-mode polarization from inhomogeneous reionization with 
various parameters for the bubble size distribution (models 12-17 in 
Table~\ref{tabrsig}). The three curves with 
$\lpeak\approx 40$ have $R_{\rm eff}=500$ Mpc (\emph{solid, thin}), and the three  
with $\lpeak\approx 400$ have $R_{\rm eff}=50$ Mpc (\emph{solid, thick}), 
with fixed $R_{\rm eff}$ during reionization. 
For each value of $R_{\rm eff}$, from the widest spectrum to the narrowest 
(lowest to highest peak amplitude), $\slnr=1$, 0.5, and 0.2 respectively. 
For all models, the other parameters are $z_i=30$, $z_f=6$, and $\tau_*=0.12$, 
with a constant ionization fraction during reionization. 
The dashed curve 
shows the predicted $B$ polarization from gravitational lensing of $E$-modes for 
our assumed cosmological parameters.
\vskip 0.5cm}
\label{clbreff}
\end{figure}

The reionization models in Table~\ref{tabrsig} and Figure~\ref{clbreff} 
show the dependence of 
$\lpeak$ and $\cclbpeak$ on \ion{H}{2} bubble size. 
Larger values of $R_{\rm eff}$ lead to a smaller multipole
for the $B$-mode peak (see equation~(\ref{lpkrsig}))
\begin{equation}
\lpeak\approx 100 \left(\frac{ 200~{\rm Mpc}}{R_{\rm eff}}\right).
\label{lpeakscaling}
\end{equation}
We can see the relation between $R_{\rm eff}$ and 
$\lpeak$ in the models in Table~\ref{tabrsig}. 
For example, models 12-14 have different values of $\slnr$ but the same 
$R_{\rm eff}$, and $\lpeak$ is the same to within about 10\%.
As $R_{\rm eff}$ increases, not only does $\lpeak$ decrease but also 
the amplitude of the peak increases.  For a fixed
$\slnr$ it scales as
$\cclbpeak\propto R_{\rm eff}$.  

The dependence of the amplitude on the width $\slnr$ at fixed 
$R_{\rm eff}$ is much weaker.    Note that
the $B$-mode peak amplitude for the dominant shot noise contributions can be approximated as
\begin{equation}
\cclbpeak\sim R_0^3 \langle W_R^2(k_{\rm eff})\rangle\sim e^{-7.5\slnr^2}
\langle W_R^2(k_{\rm eff})\rangle \,, \label{clbpkrsig}
\end{equation}
for fixed $R_{\rm eff}$.
Defining
\begin{equation}
\kappa_{\rm eff}\equiv k_{\rm eff}\bar{R}=\left(\frac{9}{2}\right)^{1/4}e^{-4\slnr^2},
\end{equation}
equation~(\ref{w1bln}) shows that $\langle W_R^2(k_{\rm eff})\rangle$ is a 
function of $\slnr$ only.
Evaluating the integral in equation~(\ref{w1bln}), we find that 
$e^{-7.5\slnr^2}\langle W_R^2(k_{\rm eff})\rangle$ is maximized as 
$\slnr\rightarrow 0$ and decreases as $\slnr$ increases, suggesting 
that $\cclbpeak$ has the same behavior.
This happens because the spectrum is more sharply peaked as 
$\slnr\rightarrow 0$, which follows from the fact that the window 
functions $\wrkone$ for smaller values of $\slnr$ bend more 
sharply at $k=k_{\rm eff}$ (see Figure~\ref{win1b}). 
This maximum amplitude is estimated to be approximately 3 times larger 
than for the same model with $\slnr=1$ (see Figure~\ref{clbreff}).

The models in Table~\ref{tabrsig} also show this trend with $\slnr$, and more generally
for the range of $\slnr$ 
explored in our models, the scaling of the peak amplitude with 
$\slnr$ at fixed $\lpeak$ shows good quantitative agreement with 
equation~(\ref{clbpkrsig}).

Models 12-17 and the approximate scaling of $\lpeak$ and $\cclbpeak$ 
with $R_{\rm eff}$  assume constant $R_{\rm eff}$ throughout 
reionization, but the same general results apply when 
$R_{\rm eff}$ is changing.  
The only difference is that the constant 
radius must be replaced in the scaling relations by an 
effective radius averaged over the duration of reionization, weighted 
by the redshift-dependent factors that appear in the integral 
for $\cclb$.  Models with fixed-$\bar n_b$ in which the radius grows 
with time 
are shown in Table~\ref{tabxe} (18-21). The $\lpeak$ values for 
these models indicate that the redshift-averaged radius for these models is 
$R_{\rm eff}\sim 300$ Mpc ($R_0\sim 25$ Mpc).

For a slowly varying $R_{\rm eff}$ and $x_e$, equation~(\ref{clbapprox})
implies that half the shot noise contributions come from
$0.8 <   (1+z)/(1+z_i)  < 1$.   The $B$-mode amplitude roughly follows
the scalings of the constant $R_{\rm eff}$ model with $R_{\rm eff}$ 
evaluated at  $z= 0.8(1+z_i) -1$.    This expectation is borne out by models 5 and
19 (see upper panel of Figure~\ref{xezfig}).  More generally for models where $R_{\rm eff}$
decreases with redshift from its final value, as in the fixed-$\bar n_b$ models, 
the $B$-mode peak decreases in amplitude and shifts to higher multipoles.

\subsection{Reionization Redshift and Total Optical Depth}

\begin{figure}
\epsscale{1.0}
\centerline{\psfig{file=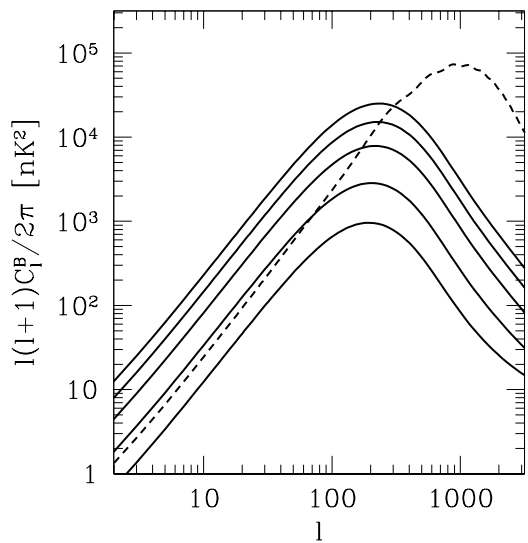, width=3.0in}}
\caption{$B$-mode polarization spectra from inhomogeneous reionization, 
varying the duration of reionization as in models 22-26 in 
Table~\ref{tabz} (\emph{solid curves}).  From top 
to bottom, the epoch of reionization begins at $z_i=50$, 40, 30, 20, and 
15. For all models, the other parameters are $z_f=6$, $\tau_*=0.12$, 
$\slnr=0.5$, and $R_{\rm eff}=93$ Mpc ($R_0=50$ Mpc). The models have 
constant $x_e$ and $R_{\rm eff}$ during reionization. The dashed curve 
shows the predicted $B$-modes from lensing.
\vskip 0.25cm}
\label{clbzi}
\end{figure}

Table~\ref{tabz} and Figure~\ref{clbzi} show models 
with different values of $z_i$, $z_f$, and $\tau_*$.
The scalings with these parameters can be understood from the shot noise
approximation in equation~(\ref{clbapprox})
\begin{equation}
\cclb\propto x_e(1-x_e)[(1+z_i)^{7/2}-(1+z_f)^{7/2}]
\label{clbshot}
\end{equation}
if $x_e$ is constant during reionization, with $x_e(\tau_{\rm rei},z_i,z_f)$ 
given by equation~(\ref{constxe}).
Note that the available optical depth during the inhomogeneous period
$\tau_{\rm rei}$ sets the ionization fraction, 
not the total optical depth $\tau_*$, and 
$\tau_{\rm rei}$ is about a factor of 4 larger when $\tau_*=0.12$ than 
when $\tau_*=0.06$ if $z_f=6$. The ionization fraction between these two 
cases then also differ by roughly a factor of 4, but this translates 
into a somewhat smaller difference in the $B$-mode amplitudes since 
$\cclb$ depends on $x_e(1-x_e)$ and not simply on $x_e$ alone.

The models in Table~\ref{tabz} obey this scaling to good approximation.
The largest polarization signals are obtained for models with high $z_i \gg z_f$ 
where $\cclb \propto (1+z_{i})^{2}$ (model 26, see also Figure~\ref{clbzi}) and external bounds on this parameter are the most
important to limit $B$-mode contributions (see Section~\ref{bounds}).    For less extreme cases, the scaling
of equation~(\ref{clbshot}) still holds.

Models 27 and 28 in Table~\ref{tabz} show that ending reionization earlier 
leads to smaller $B$-modes from inhomogeneous reionization. There are 
two reasons for this: the epoch of reionization is shorter, which decreases 
the term in square brackets in equation~(\ref{clbshot}); and increasing 
$z_f$ makes the available optical depth $\tau_{\rm rei}$ smaller, which gives a smaller value for $x_e$. 
(Note that this second effect could instead act to increase the size of 
the $B$ polarization if $x_e>0.5$.) One can check that  
equation~(\ref{clbshot}) provides a good description of the scaling of 
$B$-mode amplitudes with $z_f$ to within a few percent accuracy.

Current observational bounds on the total optical depth to 
reionization allow for a range of values of $\tau_*$ that can 
have a significant effect on the amplitude of the $B$-mode spectrum. 
Comparing models 26 and 30 in Table~\ref{tabz}, we see that increasing 
$\tau_*$ by a factor of 2 makes the amplitude larger by a factor of 
about 3.7.  For the shorter epoch of reionization in models 22 and 29, 
a factor of 2 increase in $\tau_*$ amplifies the polarization by 
a factor of 1.6.
Again, these results are well described by equation~(\ref{clbshot}) 
with $x_e$ determined by $\tau_*$.

In summary, 
the $B$-mode amplitude is maximized by making the effective bubble radius 
$R_{\rm eff}$ match the multipole of interest, making   
the duration of the epoch 
of reionization as long as possible, and taking the total optical depth 
to be as large as possible.
Narrow distributions generate the most signal but the dependence is weak
out to 
$\slnr \lesssim 1$.

\begin{center}
\begin{table}
\caption{Multipoles and amplitudes of $B$ polarization spectra peaks 
for reionization models varying $z_i$, $z_f$, and $\tau_*$.}
\begin{tabular}{|c|ccccccc|cc|}
 & & & & & fixed & & $R_{{\rm eff},f}$ & & $(\cclb)_{{\rm peak}}$ \\
Model & $\tau_*$ & $z_i$ & $z_f$ & $x_e(z)$ & $R_{\rm eff}/\bar{n}_b$ & $\slnr$ & [Mpc] & $\lpeak$ & [nK$^2$] \\
\hline
22 & 0.12 & 15 & 6 & C & both & 0.5 & 93 & 194 & 962 \\
23 & 0.12 & 20 & 6 & C & both & 0.5 & 93 & 204 & 2853 \\
24 & 0.12 & 30 & 6 & C & both & 0.5 & 93 & 217 & 7872 \\
25 & 0.12 & 40 & 6 & C & both & 0.5 & 93 & 226 & 15154 \\
26 & 0.12 & 50 & 6 & C & both & 0.5 & 93 & 232 & 25108 \\
27 & 0.12 & 30 & 8 & C & both & 0.5 & 93 & 218 & 6272 \\
28 & 0.12 & 30 & 10 & C & both & 0.5 & 93 & 218 & 4025 \\
29 & 0.06 & 15 & 6 & C & both & 0.5 & 93 & 194 & 614 \\
30 & 0.06 & 50 & 6 & C & both & 0.5 & 93 & 232 & 6792 \\
\end{tabular}
\vskip 0.75cm
\label{tabz}
\end{table}
\end{center}

\begin{figure}
\epsscale{1.0}
\centerline{\psfig{file=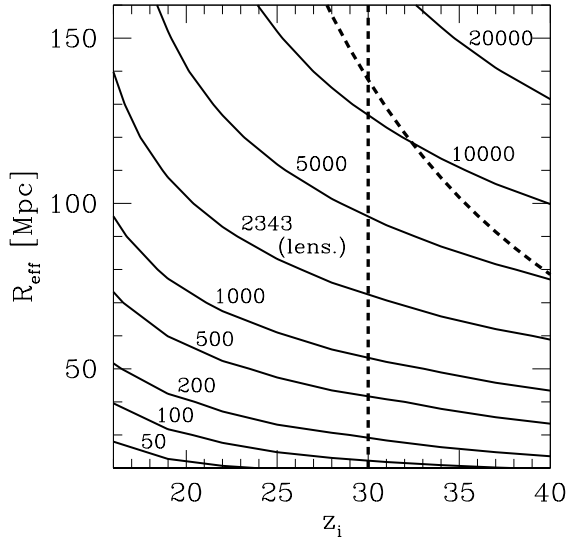, width=3.0in}}
\caption{Contours of $\cclb$ at $\ell=100$ in the 
$z_i-R_{\rm eff}$ plane. Other reionization parameters are set to 
$z_f=6$, $\tau_*=0.12$, and 
$\slnr=0.2$, and both $x_e$ and $R_{\rm eff}$ are constant during 
reionization in these models. The units of the labeled contours are 
nK$^2$. The contour at 2343 nK$^2$ marks the power in the lensing 
$B$-modes at $\ell=100$. 
The dashed curve in the upper right corner 
shows the comoving radius of a bubble at 
redshift $z_i$ such that the optical depth from that single bubble 
along a line of sight passing through the center of the bubble 
would be equal to the total optical depth, $\tau_*=0.12$. 
The vertical dashed line is at the approximate upper limit on $z_i$ 
from the observed shape of the $E$-mode reionization peak.
\vskip 0.25cm}
\label{contourfig}
\end{figure}

\section{Bounds on Reionization Parameters}
\label{bounds}

From the previous sections we conclude that the parameters that have the 
greatest effect on the $B$-mode signal from inhomogeneous 
reionization are $R_{\rm eff}$ and $z_i$. Other parameters can affect 
the amplitude, but the degree to which they can do so is limited 
either on theoretical ($x_e(z)$, $\slnr$) or observational grounds ($\tau_*$, $z_f$). 
Figure~\ref{contourfig} shows contours of $\cclb$ at 
$\ell=100$ in the $z_i-R_{\rm eff}$ plane, with $z_f=6$ and $\tau_*=0.12$ 
chosen to maximize the amplitude. We take $\slnr=0.2$ since the amplitude 
at the peak is maximized at small values of $\slnr$. As Figure~\ref{clbreff} 
shows, the amplitude on the tails of the spectrum may be higher for 
larger values of $\slnr$, but the difference is no more than a factor of a 
few.

The $B$-mode amplitude continues to grow as one increases $R_{\rm eff}$ 
or $z_i$, so the maximal contributions are determined by observational or 
theoretical constraints on these quantities.

\subsection{Observational Constraints}

The starting redshift of reionization is limited by the shape and amplitude of
the $E$-mode peak detected in WMAP \citep{page_etal06}.
The $E$-mode peak scale is determined by the horizon scale during reionization
and an early period of reionization shifts the peak to higher multipoles than observed.
For example, consider a model with fixed optical depth $\tau_{*}=0.09$ and constant ionization
fraction out to $z_{i}=30$. In this model, the suppression of power at $\ell=3$, where the weight of
the detection lies, is a factor of
$2$ larger than in an instantaneous reionization model with $\tau_{*}=0.06$, while the
enhancement of power at $\ell=10$ is a factor of $4$ larger than in a model with $\tau_{*}=0.12$.
This is consistent with the findings of \citet{spergel_etal06} for constant $x_{e}$ models
out to $z_{i}=25$ and implies that $z_{i}<30$
at approximately the $2\sigma$ level.  We show this bound as the vertical dashed line
in Figure~\ref{contourfig}.

With this bound, models that have \emph{B}-mode power in excess of the lensing 
signal at $\ell =100$ have $R_{\rm eff}\gtrsim 70$~Mpc.  
Bubbles with large radii
at high redshift also contain a large optical depth.
The optical depth from a fully ionized bubble at redshift $z$ with radius $R$ along a 
line of sight that passes through the center of the bubble is
\begin{equation}
\tau_b=\sigma_T n_{p,0}(1+z)^2 2R.
\end{equation}
This single bubble optical depth can exceed the total $\tau_{*}$ if such bubbles
have sufficiently low number densities. Despite being rare, they still produce an observable
effect on the temperature power spectrum.  In the direction of such a bubble, the
temperature power spectrum will be suppressed by an additional factor of $e^{-2 \tau_{b}}$.
These bubbles would produce a power spectrum of the acoustic peaks which varied
from field to field and experiment to experiment.  No such effect has been seen
and the differences between measurements can be attributed to the $\sim 5$-$10\%$
calibration errors. 

Such an effect is also limited by the field-to-field variance of the WMAP power spectrum.
 \citet{hansen_etal04} 
found no evidence for excess variance beyond the
10\% sample variance of the amplitude of the first peak
 when measured in different $19\degr$-radius 
patches on the sky.

As a conservative limit, we require that the optical depth from a 
single bubble should be smaller than the total optical depth to 
reionization $\tau_{*}=0.12$ or a $20\%$ variation in the power in the acoustic peaks on and off
of a bubble. For the radius, this constraint becomes
\begin{equation}
R\lesssim 140~{\rm Mpc}\left(\frac{\tau_b}{0.12}\right)\left( \frac{1+z}{31}\right)^{-2}.\label{rbound}
\end{equation}
Since the typical size of a bubble that contributes to the $B$ polarization 
from inhomogeneous reionization is $R_{\rm eff}$, we apply this 
constraint to the effective radius. 
This bound is shown for $\tau_b=0.12$ as the dashed curve in 
Figure~\ref{contourfig}.  Combined with equation~(\ref{lpeakscaling}), this also implies a bound
on the peak multipole of
\begin{equation}
\lpeak \gtrsim 140 \left( \frac{\tau_b}{0.12} \right)^{-1} \left( \frac{1+z}{31}\right)^{2} \,.
\end{equation}
Therefore even if the bound on $z_i$ is relaxed the \emph{B}-mode contributions at $\ell \sim 100$
would be in
the steeply falling white noise tail of the spectrum.  Correspondingly the maximal contributions
for models in Figure~\ref{contourfig} that satisfy this constraint occur for $z_i \lesssim 25$.

This constraint on $R_{\rm eff}$ scales linearly 
with the optical depth $\tau_b$, so future work similar to that 
of~\citet{hansen_etal04} could potentially strengthen  this bound.  For example,
a bound of $\tau_{b}<0.06$ would exclude models with \emph{B}-modes above
the lensing signal at $z_{i}=30$.

These bounds can potentially be evaded if the width of the distribution $\slnr$
is so wide that bubbles with $R \sim R_{\rm eff}$ represent such a small fraction
of the total that a typical line of sight would never intersect such a bubble.
However, if such bubbles are the source of the \emph{B}-modes,
then there would be a large fraction of sky that would be free of this signal.  These clean
regions of sky could then be used to measure gravitational wave \emph{B}-modes.
To quantify these considerations, we calculate the covering fraction of the sky
in bubbles with $R \sim R_{\rm eff}$ as a function of $\slnr$.

We define the covering fraction as follows. Consider a radius $R_p$ such 
that $p$\% of the power in the one-bubble shot noise term comes from 
bubbles with radii $R>R_p$. The median bubble radius that contributes to $B$-modes is 
$R_{50}$, which is within 5-20\% of $R_{\rm eff}$ for the range of 
$\slnr$ that we consider. We select bubbles with $R\sim R_{\rm eff}$ that 
contribute half the total power by requiring bubble radii to satisfy 
$R_{75}<R<R_{25}$. For $\slnr=0.2$, $R_{75}=0.9R_{\rm eff}$ and 
$R_{25}=1.2R_{\rm eff}$; for $\slnr=1$, $R_{75}=0.8R_{\rm eff}$ and 
$R_{25}=1.6R_{\rm eff}$.

We also choose a representative range in redshift for the covering fraction. 
Since most of the polarization signal comes from high-redshift bubbles, 
we select redshifts $z_i-\Delta z<z<z_i$ at which bubbles contribute 
half the total $B$-mode power. In the reionization models we 
consider, $\Delta z$ varies between about $0.15(1+z_i)$ and $0.25(1+z_i)$, so 
we approximate the width in redshift as $\Delta z\approx 0.2(1+z_i)$ (see Section~\ref{sizedist}). 
Since bubbles with radii between $R_{75}$ and 
$R_{25}$ contribute half the total power at any given redshift, combining 
the restrictions on bubble radius and redshift leaves a sample of 
bubbles that produce a quarter of the $B$-mode power from 
inhomogeneous reionization.

\begin{figure}
\epsscale{1.0}
\centerline{\psfig{file=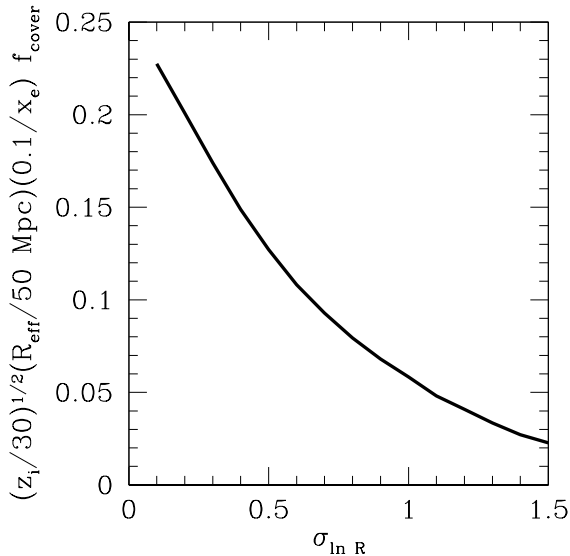, width=3.0in}}
\caption{The covering fraction of bubbles with radii $R\sim R_{\rm eff}$ at 
redshifts $z\lesssim z_i$ that are responsible for about 1/4 of the total 
shot noise power. The covering fraction depends on $R_{\rm eff}$, $z_i$, and 
$x_e$ as indicated in the axis label.
\vskip 1.0cm}
\label{fcover}
\end{figure}

In a spherical shell at redshift $z_i$ with width $\Delta z$, the number 
of bubbles per solid angle with radius $R$ is
\begin{equation}
\frac{dN}{dRd\Omega}=P(R)\bar{n}_b D_A^2(z_i)\frac{\Delta z}{H(z_i)},
\end{equation}
and the solid angle subtended by a bubble of radius $R$ at distance 
$D_A$ is $\pi R^2/D_A^2$. Then the covering fraction of bubbles that 
meet the criteria for radius and redshift is
\begin{eqnarray}
f_{\rm cover}&=&\int_{R_{75}}^{R_{25}}dR
\frac{dN}{dRd\Omega}\frac{\pi R^2}{D_A^2}\label{fcovereqn}\\
&=&\frac{3(1+z_i)x_e}{40H(z_i)R_{\rm eff}}e^{3\slnr^2/2}\nonumber\\
&\times&\left[\erf\left(\frac{2\slnr^2+\ln r_{25}}
{\sqrt{2}\slnr}\right)-\erf\left(\frac{2\slnr^2+\ln r_{75}}
{\sqrt{2}\slnr}\right)\right],\nonumber
\end{eqnarray}
where we define $r_p\equiv R_p/R_{\rm eff}$ and assume $\Delta z=0.2(1+z_i)$ 
and $x_e\ll 1$. Note that 
this is likely to be an overestimate as the covering fraction approaches 
unity since we assume that the bubbles are not overlapping along lines 
of sight.

The covering fraction for $z_i=30$, $R_{\rm eff}=50$ Mpc, and $x_e=0.1$ 
is shown in Figure~\ref{fcover}. It becomes negligible for distributions with
$\slnr \gtrsim 1$.  The covering fraction decreases for 
larger $z_i$, larger $R_{\rm eff}$, or smaller $x_e$.

\subsection{Theoretical Expectations}

Observational constraints still allow models with large \emph{B}-mode
contributions, approaching and in some cases exceeding the gravitational
lens contributions (see Figure~\ref{contourfig}).  However they require $R_{\rm eff} \gtrsim 100$~Mpc
at $z_i \sim 20$-$30$. These models are currently disfavored
theoretically as plausible reionization scenarios.

Normal population II star formation is unlikely to be sufficient to
cause significant ionization above $z \sim 15$ \citep[see][and references therein]{barkana_loeb01}.  One possibility for obtaining substantial ionization at high redshift is a 
``double reionization'' model where a change in the ionizing sources to say  metal free
population III stars causes non-monotonic behavior in the ionization fraction
\citep{Cen02,WyiLoe02}.  Such models were studied extensively following
the indications from the first year WMAP data of a high total optical depth.

Recently, \citet{furlanetto_loeb05} have argued that such models are unlikely to 
cause a non-monotonic behavior.  
This would rule out 
models that have significant ionization at high redshifts 
compensated by a low value of $x_e$ at intermediate redshifts to keep 
the total optical depth within the observed limits, such as the 
maximal $B$-mode ionization histories we consider.   However, 
\citet{furlanetto_loeb05} find that models with extended periods of 
ionization in which $x_e$ is significant over a wide range of redshifts out to $z_i \lesssim 20$ are not difficult to find.  Recall that such models
would produce \emph{B}-modes that are within a factor of a few of maximal at
a given $z_i$.  However given the low $\sigma_{8}$ currently indicated by WMAP
\citep{spergel_etal06}, extended reionization at high redshift becomes even
more difficult to achieve~\citep{alvarez_etal06}.  Recent simulations have nonetheless
found that substantial ionization at $z=15$-$20$ is still possible with small-mass
sources~\citep{iliev_etal06}. 

Even for $z_i \lesssim 20$ substantial \emph{B}-modes are possible but only if the
effective bubble radius $R_{\rm eff} \gtrsim 100$ Mpc.  
The clustering of sources could create \ion{H}{2} regions with comoving 
radii of 10-30 Mpc near the end of reionization, larger than might 
be expected for a bubble around a single source 
galaxy~\citep{wyithe_loeb05}.   Note also that $R_{\rm eff}$ is
much greater than the typical radius $\bar R$ of a bubble if the distribution is wide.
However, dense absorbers such as 
Lyman limit systems that are found to be associated with galaxies 
in simulations could restrict the size of bubbles~\citep{gnedin_fan06,kohler_gnedin06}.
Furthermore, near the beginning of reionization the existing bubbles should
be much smaller in radius. 

\begin{figure}
\epsscale{1.0}
\centerline{\psfig{file=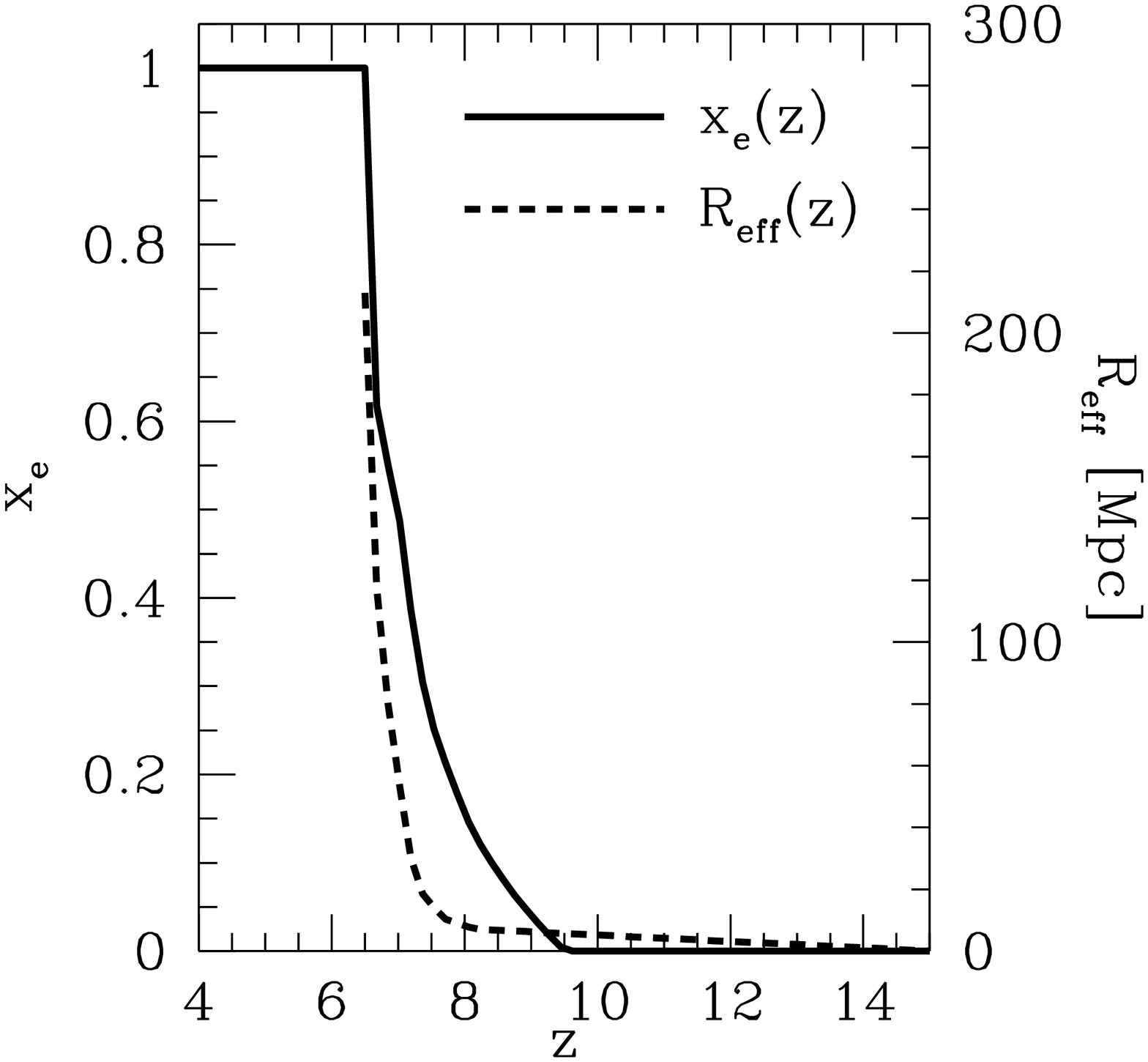, width=3.0in}}
\centerline{\psfig{file=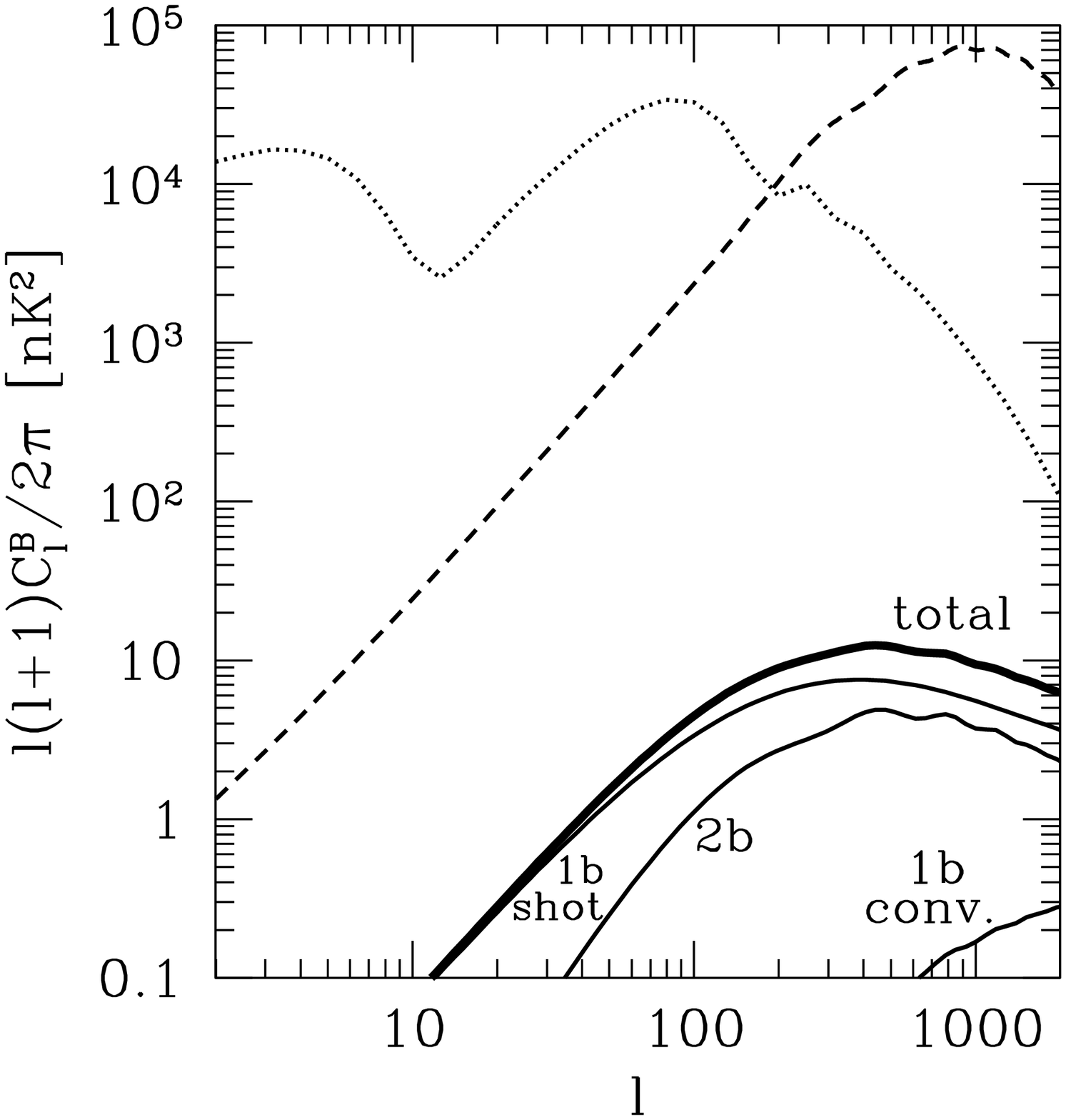, width=3.0in}}
\caption{Ionization history and bubble size (\emph{upper panel}) and $B$ 
polarization from inhomogeneous reionization (\emph{lower panel}) for a 
model intended to approximate the results of the reionization 
simulations of \citet{zahn_etal06}. 
The width of the bubble size 
distribution is chosen to be $\slnr=0.7$. As in Figure~\ref{clbcomps}, 
the lensing $B$-modes (\emph{dashed}) and gravitational wave $B$-modes 
(\emph{dotted}) are shown for comparison. The components of the 
inhomogeneous reionization power spectrum are also plotted as in 
Figure~\ref{clbcomps}. The total optical depth for this 
model is $\tau_*=0.055$ and the duration of reionization is fairly short 
so this model can be taken as nearly
the minimal \emph{B}-modes from inhomogeneous reionization.
\vskip 1.0cm}
\label{zahnmodel}
\end{figure}

To make these considerations concrete, we examine 
the analytic model of \citet{furlanetto_etal04} which makes 
specific predictions for the bubble distribution when
coupled with the
simulations of \citet{zahn_etal06} for the ionization history.
 Here reionization occurs fairly promptly and
at low redshift.
In this model,
the peak radius of the size distribution
is of order $\bar{R}\sim 1$ Mpc (comoving) 
early in reionization and increases to $\bar{R}\sim 10$-100 Mpc by 
the end of reionization~\citep{furlanetto_etal05}. 
The shape of the distribution shows some skewness toward smaller bubbles
but can be crudely modelled by a log normal 
with a width that increases with redshift from $\slnr\sim 0.5$ 
to $\slnr\sim 1$.    For simplicity
we approximate this as a constant $\slnr =0.7$.
Both the ionization history and the bubble distribution in this
model act to lower the \emph{B}-mode contributions as shown in
Figure~\ref{zahnmodel}.
They are several orders of magnitude below the lensing $B$-modes and would
be relevant only if the polarization field were cleaned of lensing on the whole 
sky.    
This model can also be considered near the minimal expected contribution in
that the total optical depth is quite low ($\tau_*=0.055$) and the duration of
reionization quite short.

The model in Figure~\ref{zahnmodel} illustrates some exceptions to 
our earlier conclusions. Because $x_e$ and $R_{\rm eff}$ change 
rapidly with redshift, the redshift-averaged effective radius is not 
necessarily well approximated by $R_{\rm eff}(0.8z_i)$ as described in 
Section~\ref{sizedist}. The location of the peak at $\lpeak\sim 400$ 
indicates that the main contribution is from bubbles with radii 
$R\sim 50$ Mpc, and $R_{\rm eff}$ does not reach this size until 
around $z=7$. Related to this is the fact that in this model, 
nearly all of the $B$-mode power comes from bubbles at $z\lesssim 7$, 
at the very end of reionization. In the models we examined where 
reionization is a more gradual process, a large fraction of the 
power comes instead from the highest-redshift bubbles.

Our model assumes that high-density regions that host ionizing sources 
are ionized before areas with lower density, consistent with recent 
simulations of reionization \citep[e.g.,][]{Cen02,Soketal03,iliev_etal06,gnedin_fan06,zahn_etal06}.
Models in which reionization proceeds in the reverse order have also 
been proposed, and such a scenario could produce different $B$-mode 
spectra from those studied here~\citep{mhr00}. 
If reionization does begin in voids, we 
would expect the two-bubble contribution to the $B$ polarization to 
be affected the most. As long as the \ion{H}{2} regions in such a scenario 
could still be characterized as roughly spherical ionized regions 
with a size distribution that is approximately log normal, our 
model should still be sufficient to describe the shot noise contribution 
to the $B$-modes.

\section{Discussion}
\label{discussion}

We have presented a comprehensive study of the generation of \emph{B}-mode
polarization by inhomogeneous reionization with an emphasis on the phenomenological
properties required to generate a substantial contamination for degree scale
gravitational wave studies.  We base our study on a general
parameterization of inhomogeneous reionization in terms of three functions of
redshift: the mean ionization fraction, the effective bubble radius, and the width
of a log normal radius distribution.

 We find that these \emph{B}-modes are
only important when the ionized bubbles are sufficiently rare that the power spectrum
is dominated by the shot noise of the bubbles.  In such models, the \emph{B}-modes
are maximized by taking an effective bubble radius as large as possible, up to a few
hundred Mpc, at as high a redshift as possible for a total duration of reionization that
is as long as possible, while still remaining consistent with constraints on the total optical depth.  
The width of the bubble distribution is less critical but a wider
distribution can make the effective bubble radius much larger than the typical bubble.
The details of the ionization history are also less relevant so long as a finite ionization
fraction remains at high redshift.  We provide useful scaling relations for estimating
the \emph{B}-mode contributions for a wide variety of scenarios.

These conditions are constrained both observationally and by theoretical expectations.
A substantial ionization fraction at high redshift would violate the shape of the
observed \emph{E}-mode power spectrum which sets a maximum redshift of $z \sim 30$.
A large bubble radius would produce an large optical depth through its center
and hence inhomogeneous  obscuration of the acoustic peaks.  Combined these 
conditions still allow power at $\ell=100$ that is $\sim 5$ times the power produced
by gravitational lensing or $\sim 0.1~\mu$K in the polarization field.    These amplitudes
can be reached for example by $\sim 140$ Mpc bubbles at $z \sim 30$.   This level
is also comparable to the {\it maximum} allowed \emph{B}-modes from gravitational
waves at $\ell=100$.  On the other
hand, not even these models
 would predict \emph{B}-modes at $\ell \lesssim 10$ that would mask
the maximal gravitational wave signal from reionization.

Current theoretical models and simulations
 would not predict the existence of such large ionization bubbles at such high redshifts.
 In these models bubbles grow to a few tens of Mpc only toward the end of a fairly 
 prompt reionization. Nonetheless, 
 even for conditions where the \emph{B}-modes do not exceed the lensing signal,
they can still be the leading order cosmological contaminant to the gravitational wave
signal.   

Even without de-lensing of the polarization, the well-defined power spectrum from lensing
allows its contributions to be statistically subtracted,  ultimately to the few percent level
with a full-sky cosmic variance limited experiment.
The uncertain form of the inhomogeneous reionization \emph{B}-modes does not permit
statistical subtraction.  With $\tau_*=0.12$, $\sim 20$ Mpc bubbles at $z \sim 15$ near the
end of reionization would produce \emph{B}-modes
that are $1\%$ of the power in lensing at $\ell =100$ or $10\%$ of the polarization
field.   Likewise with a lower total optical depth of $\tau_*=0.06$, $\sim 40$ Mpc bubbles at
$z \sim 10$ would produce the same effect. 
Such contributions would also be relevant 
for the percent precision measurements of the lensing \emph{B}-modes on smaller
scales expected from the next generation of ground based experiments.

\acknowledgements {\it Acknowledgments}: We thank S. DeDeo, O. Dor\'{e}, 
 N. Gnedin, D. Huterer, K.M. Smith, Y.S. Song, 
X. Wang and B. Winstein for useful conversations.  MJM was supported by a 
National Science Foundation Graduate Research Fellowship. 
WH was supported by the KICP
through the grant NSF PHY-0114422, the DOE through 
contract DE-FG02-90ER-40560 and the David and
Lucile Packard Foundation.

\bibliographystyle{apj_hyperref}

\end{document}